\documentclass[apj]{emulateapj}
\usepackage{graphicx}  
\usepackage{amsmath}
\usepackage{bm}
\usepackage{natbib}
\newcommand{\ud}{\,\mathrm{d}}

\newcommand{\rex}{\, R_\mathrm{ex}}

\newcommand{\rin}{\, R_\mathrm{in}}
\newcommand{\zsp}{\, Z_\mathrm{sp}}
\newcommand{\vex}{\, V_\mathrm{ex}}
\newcommand{\vin}{\, V_\mathrm{in}}
\newcommand{\circf}{\, \mathrm{circ}}
\newcommand{\xringf}{\, \mathrm{xring}}
\newcommand{\gax}{\, g_\mathrm{ax}}
\newcommand{\bp}{\,\mathbf{p}}
\newcommand{\bu}{\,{\bf u}}
\newcommand{\bx}{\,\mathbf{x}}

\begin{document}
\DeclareGraphicsExtensions{.png, .eps, .jpeg, .giff, .pdf}
\slugcomment{Draft text, {\today}}
\journalinfo{Target: AstroPhysical Journal}
\title{Reconstruction of Static Black Hole Images Using Simple Geometric Forms}
%
%
%


\author{Leonid Benkevitch, Kazunori Akiyama, Rusen Lu, Shepherd Doeleman, Vincent Fish  \emph{et al.}}
\affil{MIT Haystack Observatory, Westford MA}
\email{benkev@haystack.mit.edu}

\begin{abstract} \small 
General Relativity predicts that the emission close to a black hole must be lensed by its strong gravitational field, illuminating the last photon orbit. This results in a dark circular area known as the black hole 'shadow'. The Event Horizon Telescope (EHT) is a (sub)mm VLBI network capable of Schwarzschild-radius resolution on Sagittarius A* (or Sgr A*), the 4 million solar mass black hole at the Galactic Center. The goals of the Sgr A* observations include resolving and measuring the details of its morphology. However, EHT data are sparse in the visibility domain, complicating reliable detailed image reconstruction. Therefore, direct pixel imaging should be complemented by other approaches. Using simulated EHT data from a black hole emission model we consider an approach to Sgr A* image reconstruction based on a simple and computationally efficient analytical model that produces images similar to the synthetic ones. The model consists of an eccentric ring with a brightness gradient and a two-dimensional Gaussian. These elemental forms have closed functional representations in the visibility domain, which lowers the computational overhead of fitting the model to the EHT observations. For model fitting we use a version of the Markov chain Monte-Carlo (MCMC) algorithm based on the Metropolis-Hastings sampler with replica exchange. Over a series of simulations we demonstrate that our model can be used for determining geometric measures of a black hole, thus providing information on the shadow size, linking General Relativity with accretion theory. \\
\end{abstract}

\section{Introduction}

The Event Horizon Telescope (EHT) is a project to observe supermassive black holes, including Sagittarius A* (Sgr A*) and M87, at an angular resolution comparable to the black hole Schwarzschild radius.  Upgrades to EHT instrumentation that are currently underway will increase the sensitivity and baseline coverage of the array, making it possible to produce images of these sources. Because of the small number of antennas and, hence, sparsity of the $uv$ coverage the reconstructed image of Sgr A* black hole and its accretion flow will have severe uncertainty or ambiguity. Therefore, additional constraints on the Sgr A* image are required. Previous observations with smaller number of the baselines allowed to make preliminary estimates of the Sgr A* event horizon size ($40-60\;\mu$as), spin (close to zero) and the viewing angle ($\sim 68^\circ$) \citep{Doeleman2008Nature, Broderick2009estim, Fish2011detect, Broderick2011lowspin, Broderick2011constr}. Other constraints, derived from theoretical considerations, describe subtler details of the Sgr A* morphology. The size and shape of the black hole shadow depend on the nearby space-time metric, and the no-hair theorem infers that the black hole space-time has the Kerr metric. \citet{Johannsen2012testsgra, Johannsen2010testqk, Johannsen2010testbhim} elaborated a framework for testing the no-hair theorem. They suggested a parameterized non-Kerr metric and considered the changes in the shadow morphology due to its deviations from the Kerr metric. 

Two major techniques can be used to analyze VLBI data: direct imaging and model fitting. In order to reconstruct the brightness image from the sparse set of visibilities, maximum entropy (MEM) or similar methods are used \citep[see, for example,][]{Narayan1986, Baron2008, Baron2010, Kluska2014, RusenLu2014}. Of all possible images corresponding to the observation data the method selects an image with the maximum entropy. The benefit of direct imaging is its model independence. However, due to the non-linearity of MEM and other similar methods (e.g. CLEAN), the relationship between visibility data errors and the noise in the resultant image is not clear. In the alternative model fitting technique, the possible brightness distribution is described by a parametric model with well-determined linear mapping on the visibility domain. Such a model can be used to calculate the expected visibility measurements. The parameters are then adjusted to minimize a criterion such as $\chi^2$. This approach allows estimation of model parameter errors arising from errors in the measured visibilities, which is a substantial advantage of the model fitting technique. However, with all its advantages, the model-fitting approach has one inherent disadvantage: to ``see" the object as its model we first must know how it ``looks" to design its model. This drawback does not devaluate the approach because both imaging and model fitting should be utilized together. Namely, the first model-independent images can be obtained via imaging. Studying the images with theoretical insight is instrumental in designing models. Thus elaborated models can be fitted to the observational data to produce much more plausible images. The main value of the fitted model is its ability to quantitatively measure the features of the observed object. 

We use a Markov Chain Monte Carlo (MCMC) method for finding best-fit model parameters along with their posterior probability distributions. Generally, the posterior distributions may be complicated---multi-modal or not bell-shaped at all. However, if the model is well designed and plausibly reflects the view of the observed object, the parameter statistics from MCMC are usually close to normal distributions with statistical moments conditioned by those in the visibility measurements. Thus the errors in estimated parameters of the model can be characterized by the standard deviations of the posterior distributions.  

By now, a variety of models of the accretion flow have been created, some based on the electron concentration and temperature profiles \citep{Yuan2003nonthermal, Broderick2006freqdepshift}, others on magnetohydrodynamics and radiative transport processes \citep{Monica2009radmod, Monica2011numeric, Fish2009constrriaf}. Direct estimation of the physical model parameters based on the observations is problematic. The existing physical models of Sgr A* are non-linear and complex. They have to take into account the effects of multiple orbiting of the photons, and the ray-tracing \citep{Psaltis2012raytr} consumes significant computational resources and time. A statistical algorithm of parameter estimation for these physical models would require an unacceptably long time. Therefore, for the Sgr A* image reconstruction a simple geometric model reflecting only the overall geometric features produced by the physical models may be preferred. 

\begin{figure*}    
  \begin{center}
    \pdfimageresolution=350
    \includegraphics{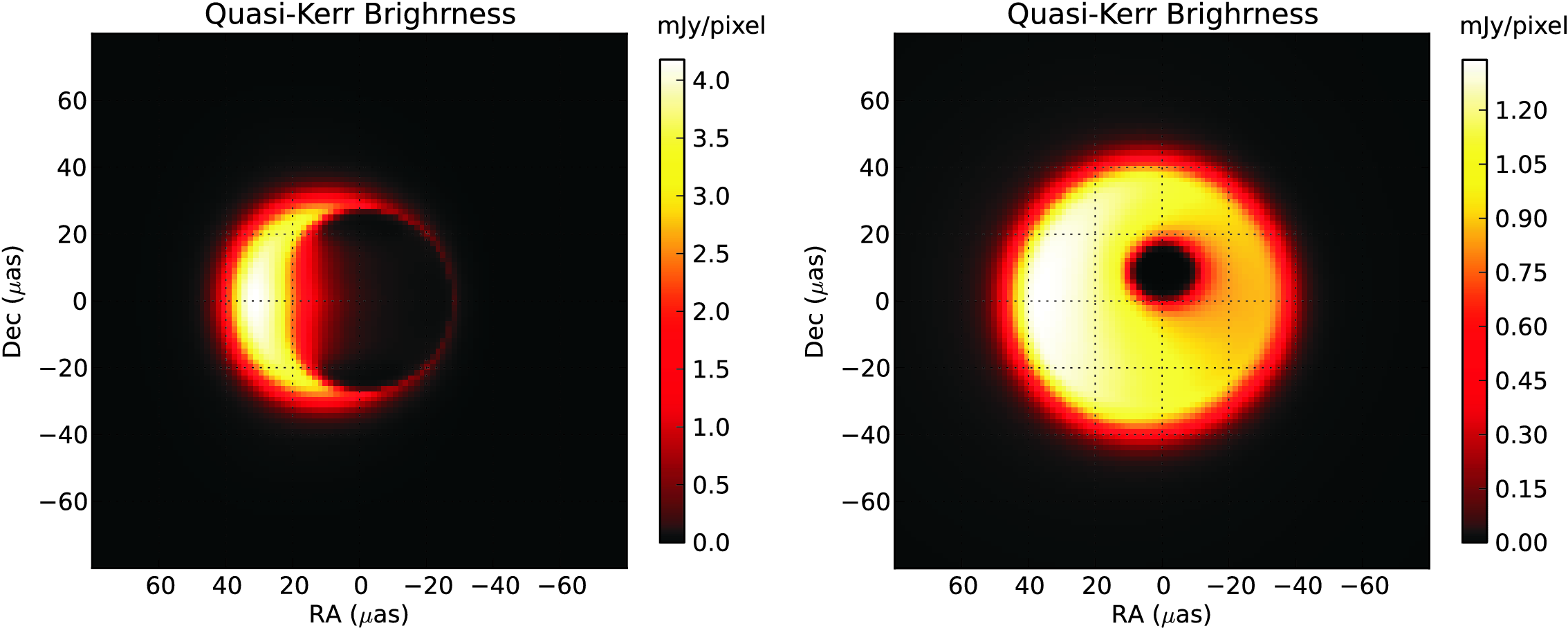}
  \end{center}
\caption{\small Simulated Quasi-Kerr Images for different inclinations of the accretion disk \citep{BJPL2013}. Left panel: the disk is close to the edge-on orientation. Right panel: the disk is close to the face-on orientation. 
\label{quasikerr_f1}}
\end{figure*}

A possible view of the black hole and its image geometry is determined by the nearby physical processes. Strong gravitational lensing makes the emission from behind the black hole appear to come from around it. Also, due to relativistic beaming the approaching side of the accretion disk appears to be many times brighter than the receding side.  If the inclination is close to $90^\circ$, the black hole looks like an eccentric ring or crescent, as in the left panel of Fig.~\ref{quasikerr_f1}. Conversely, in the case of low inclination the black hole will look like a funnel, shown in the right panel of Fig.~\ref{quasikerr_f1}. In the visibility domain these simple forms can be represented by algebraic expressions only using elementary functions to form a visibility model in the $uv$-plane that is fit to the observational data points. The $\chi^2$ distribution is calculated on the visibility magnitudes and closure phases. The inverse Fourier transform (IFT) of the best-fit model is then used to reconstruct the brightness image of the observed black hole. The analytical model must be flexible enough to resemble both states shown in Fig.~\ref{quasikerr_f1}. This significant simplification is justified by the computational speed. A similar approach has been recently used by \citet{Kamruddin_Dexter_2013}. They offered a geometric crescent model, composed of two eccentric cylinders of the opposite sign. This yields an eccentric ring crescent of uniform brightness. Our 9-parameter xringaus model provides a more detailed black hole accretion image by introducing a gradient in the crescent brightness and a two-dimensional Gaussian enhancement at the brightest part of the image. 

In the second section we describe two geometric models: the simplest ``slashed ring" and the 9-parameter ``xringaus" model. The third section is devoted to a description of the model fitting method, Markov Chain Monte Carlo with replica exchange. The fourth section describes simulations using the model and outlines the limits of the models' usability. Section five discusses our results.

\section{Slashed Ring and Gaussian Models}
The apparent shape of a black hole depends on its spin and its inclination $i^\circ$ of the accretion disk with respect to the observer. Here we use a set of Sgr A* model images created with the use of the BJPL2013 physical model developed by A.~E. Broderick, T. Johannsen, D. Psaltis, and A. Loeb \citep{BJPL2013}. Fig.~\ref{quasikerr_f1} shows two characteristic views with the edge-on (left) and almost head-on (right) positions of the accretion disk. The edge-on image shadow has the form of a Pascal lima\c con \citep{Vries2005limacon}, which can be roughly approximated by a circle. The head-on image is a non-uniformly luminous disk with a small circular shadow region. 

The images in Fig.~\ref{quasikerr_f1} can be roughly described as combinations of circular ``pillboxes" and Gaussians, all of which have closed form expressions in the visibility domain. For circularly symmetric objects we introduce their radial variables in the brightness domain as
\begin{equation}
  \label{radius_r}
  r =  \sqrt{x^2 + y^2}
\end{equation}
and in the visibility domain as
\begin{equation}
  \label{radius_rho}
  \rho =  \sqrt{u^2 + v^2}.
\end{equation}
A pillbox function $\circf(r)$ is defined by $\circf(r) = 1$ within the circle $0 \leqslant r \leqslant 1$, and is 0 otherwise. For a pillbox of radius $R$ its IFT in the visibility domain is
\begin{equation}
  \label{pillbox_vis}
  \circf\left( \frac{r}{R} \right) \leftrightharpoons \frac{R J_1(2 \pi R \rho)}{\rho}.
\end{equation}
Here and further $J_0$, $J_1$, and $J_2$ are Bessel function of the first kind. A superposition of two pillboxes, positive with the radius $\rex$ and negative with the radius $\rin$, $\rex > \rin$, makes up a luminous ring:
\begin{equation}
  \label{ring_bri}
  \mathrm{ring}\left(r \right) = 
  	  \circf\left( \frac{r}{\rex} \right) - \circf\left( \frac{r}{\rin} \right),
\end{equation}
or, in the visibility domain,
\begin{equation}
  \label{ring_vis}
  \mathrm{ring}\left(\rho \right) = 
      \frac{\rex J_1(2 \pi \rex \rho)}{\rho} - \frac{\rin J_1(2 \pi \rin \rho)}{\rho}.
\end{equation}
A ring with non-uniform brightness, specifically, with a linear gradient, can be rendered as a product of the ring function and a linear function of $x$ and $y$ coordinates (geometrically a plane). Multiplication in the brightness domain is transformed into convolution in the visibility domain. There is a theorem expressing such convolutions analytically for polynomial terms:
\begin{equation}
  \label{mult_xnfx}
  x^n f(x) \rightleftharpoons \left( \frac{i}{2\pi} \right)^n \frac{\ud^n F(s)}{\ud s^n},
\end{equation}
where $f(x) \rightleftharpoons F(s)$. In the linear case, the formula is simple:
\begin{equation}
  \label{mult_xfx}
  x f(x) \rightleftharpoons \frac{i}{2\pi} F'(s).
\end{equation}
The derivatives of Bessel functions of the first kind $J_\nu(s)$ can be expressed in terms of $J_{\nu\pm 1}(s)$ by the identities
\begin{equation}
  \label{deriv_bessel}
   \frac{\ud}{\ud s}J_\nu(s) = \frac{1}{2} \left(J_{\nu-1}(s) -  J_{\nu+1}(s) \right).
\end{equation}
The Fourier transform (FT) of a two-dimensional normalized (having integral over the $xy$ plane equal unity) Gaussian with its axes parallel to the $x$ and $y$ axes and its center at the origin is
\begin{equation}
  \label{gaussian_transform}
  \frac{1}{2 \pi \left(\sigma_a^2 + \sigma_b^2\right)} 
     e^{-\left(\frac{x^2}{2\sigma_a^2} + \frac{y^2}{2\sigma_b^2}\right)} \leftrightharpoons
         e^{-2\pi^2\left(\sigma_a^2 u^2 + \sigma_b^2 v^2\right)},
\end{equation}
where $\sigma_a$ and $\sigma_b$ are measures of the width along the $x$ and $y$ axes. Alternatively, the widths of the Gaussian can be specified in terms of the full width half maximum (FWHM),
\begin{equation}
  \label{fwhm}
  \mathrm{FWHM} = 2\sqrt{2 \ln 2}\sigma
\end{equation}
Since two-dimensional Fourier transforms obey the same rotation rules as their originals do, we can restrict the slope of the plane to the $x$ direction, and then rotate the transform to any desired angle. 

\begin{figure*}[ht!]
  \begin{center}
    \pdfimageresolution=350
    \includegraphics{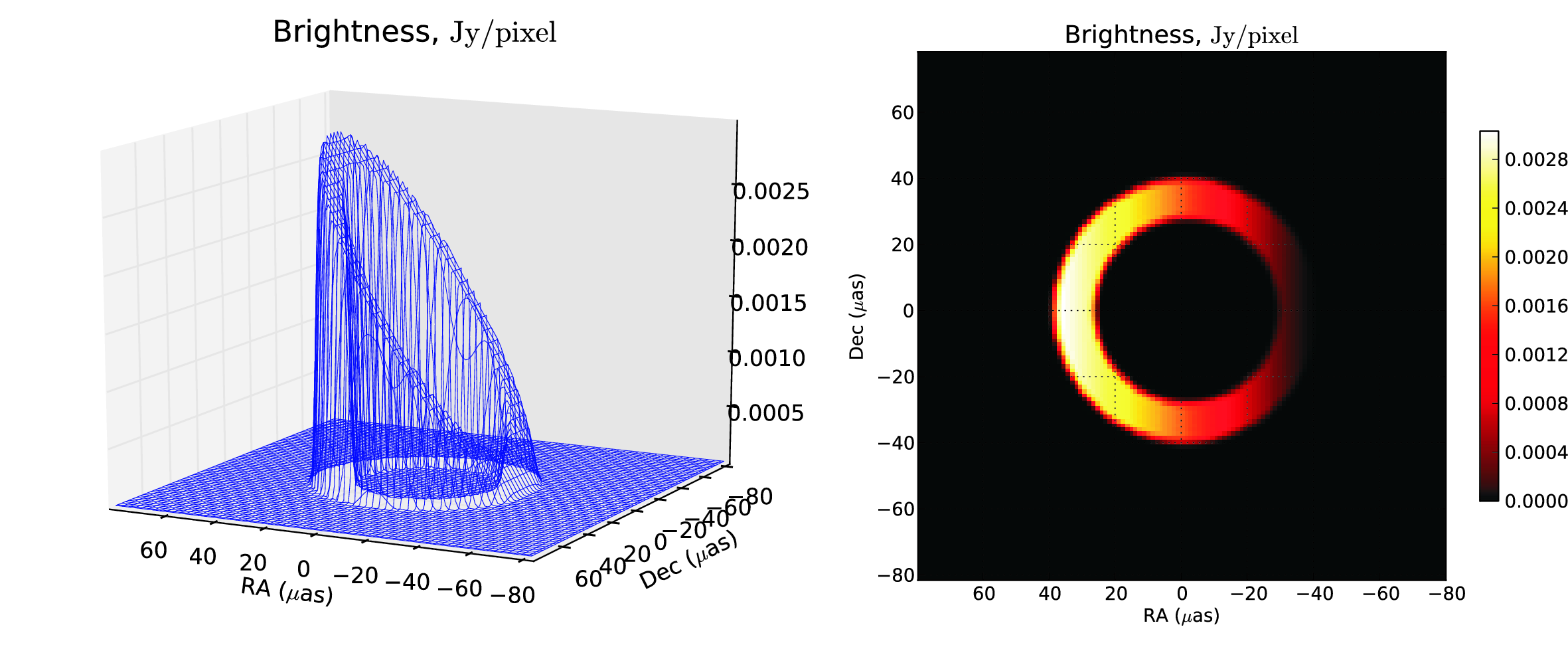}
  \end{center}
\caption{\small The ``concentric slashed ring" function as a simple model for the black hole brightness image. Left: a three-dimensional view. Right: a map view. \label{slashed_inkpot}}
\end{figure*}


\subsection{A simple ``concentric slashed ring" model}

Consider a simple case of the concentric ring from  Eqn. ~\ref{ring_vis} multiplied by the linear function 
\begin{equation}
  \label{lin_fun}
  l(x) = \frac{h}{2} \left( \frac{1}{\rex} x + 1 \right).
\end{equation}
This function represents a slanted plane with $l(-\rex) = 0$ and $l(\rex) = h$. If we regard $h$ as the maximum brightness, the product 
\begin{equation}
  \label{prod_lxring}
  b(x,y) = l(x) \, \mathrm{ring}\left(\rho, R_{ex}, R_{in} \right)
\end{equation}
will be a ring with maximum brightness $h$ at the point $(\rex,0)$ and zero brightness at $(-\rex,0)$. An example of this ``concentric slashed ring" is shown in Fig.~\ref{slashed_inkpot}. The image is obtained via the IFT of the complex visibility function $V(u,v) \leftrightharpoons b(x,y)$ back into the brightness function \ref{prod_lxring}. We consider $V(u,v)$ as a simple ``concentric slashed ring" model. It only has three parameters, $\zsp$, $\rex$, and $\rin$ and it has its maximum brightness at the point $(\rex,0)$. 

The total zero-spacing flux $\zsp$ of the model ring source is the integral over the area that encloses the ring, or its geometric volume $\zsp = \onehalf \pi(\rex^2 - \rin^2)$, 
therefore, to normalize $\zsp$ of our slashed ring to unity, $h$ must be
\begin{equation}
  \label{unity_zsp_h}
  h = \frac{2}{\pi (\rex^2 - \rin^2)}.
\end{equation}

 Since a ring is the difference between two pillboxes, we can write Eqn. \ref{prod_lxring} and its FT for each pillbox $\circf(r/R)$ separately, with $R = \rex$ or $\rin$, and then take their difference. Denote the FT of a pillbox of radius $R$ as $\circf(\rho,R)$:
 \begin{equation}
  \label{ft_of_circ}
  \circf \left(\frac{r}{R} \right) \rightleftharpoons \circf(\rho,R) 
\end{equation}
The pillbox slashed by the plane (Eqn.~\ref{lin_fun}) is
\begin{equation}
  \label{lb_bri}
  b(x,y) = \frac{h}{2} \left(1 + \frac{1}{\rex} x\right) \circf\left(\frac{r}{R} \right),
\end{equation}
and from Eqn.~\ref{mult_xfx} its FT is
\begin{equation}
  \label{lb_vis}
    V(u,v) = \frac{h}{2} \left(\circf(\rho,R) + \frac{i}{2 \pi} \frac{\ud }{\ud \rho} \circf(\rho,R)u
                         \right),
\end{equation}
where the derivative of $\circf(\rho,R)$ is
\begin{eqnarray}
  \label{F_der}
  \frac{\ud }{\ud \rho} \circf(\rho,R) &=& R\biggl[\frac{\pi R (J_0(2 \pi R \rho) - 
              J_2(2 \pi R \rho))}{\rho^2} \nonumber \\
    &-& \frac{J_1(2 \pi R \rho)}{\rho^3} \biggr] u.
\end{eqnarray}
The ``concentric slashed ring" model is thus 
\begin{equation}
  \label{slr_model}
  V(u,v) = \vex(u,v) - \vin(u,v),
\end{equation}
where $\vex(u,v)$ and $\vin(u,v)$ are computed as prescribed by Eqn.~\ref{lb_vis} with $R = \rex$ and $R = \rin$, respectively.

In order to rotate the slashed ring by an angle $\theta$ around the origin $(0,0)$ we rotate its FT image in the visibility domain using the standard coordinate transformation
\begin{equation}
  \label{rotate_uv}
  \begin{array}{llc}
  	u' &=&  u \cos(\theta) + v \sin(\theta) \\
  	v' &=& -u \sin(\theta) + v \cos(\theta).
  \end{array}
\end{equation}

\subsection{The nine-parameter ``xringaus" model}

\begin{figure*}
  \begin{center}
    \pdfimageresolution=350
    \includegraphics{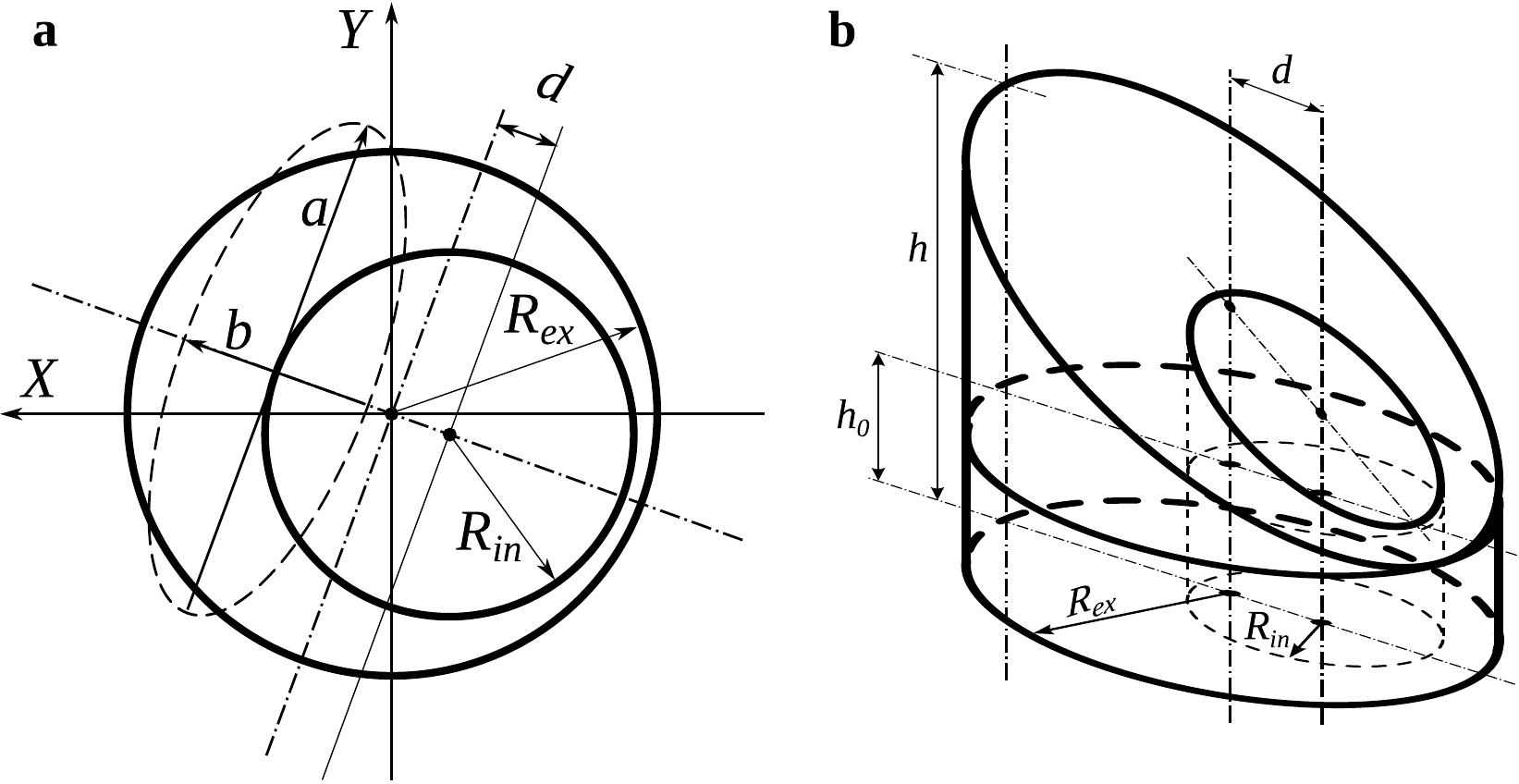}
  \end{center}
\caption{\small Geometry of the nine-parameter model as a combination of two slashed circular pillbox functions and an elliptic Gaussian. (a): View from above. The dashed ellipse indicates the FWHM of the elliptic Gaussian, with main axes $a$ and $b$. (b): axonometry to help the volume calculation.  \label{nine_param_geometry}}
\end{figure*}

This model has been designed to make the brightness images closer to the simulated quasi-Kerr images (see Fig.~\ref{quasikerr_f1}) than those of the too simple slashed ring model. We allowed internal ring displacement within the external ring by multiplying its FT by the shift operator $\exp(i 2\pi d u)$, where $d$ is the distance between the pillbox centers. Thus the ring becomes eccentric to allow arbitrary positioning of the black hole shadow. Also, an elliptical Gaussian is (optionally) added to the bright part of the ring to let the brightness outside of the ring fall off more smoothly. As a result, this model has a tuple of nine parameters: the zero-spacing flux $\zsp$, the external radius $\rex$, the internal radius $\rin$, the distance between centers of the circles $d$, the ``fading" parameter controlling the minimum brightness, the Gaussian axes $a$ and $b$, the fraction of the total flux in the Gaussian $g_q$, and the rotation angle $\theta$:
\begin{equation}
  \label{param_tuple}
  \bu = \left[ \zsp, \rex, \rin, \ d, \ f, \ a, \ b, \ g_q, \ \theta \right]
\end{equation}
Schematics of the model brightness image, detailing its components, are shown in Fig.~\ref{nine_param_geometry}. The axes are allowed to vary, while the Gaussian center has always the same position.

The model parameters must satisfy certain restrictions, like $\rex > \rin$, $d \leqslant \rex - \rin$ etc. In order to simplify the specifications of the restrictions, and hence, of the prior used in the Markov Chain Monte-Carlo fitting processes, we replace most of the parameters with their dimensionless ratios valid within the $[0,1]$ interval:
\begin{equation}
  \label{param_tuple_rel}
  \bu = \left[ \zsp, \rex, r_q, \ \epsilon, \ f, \ \gax, \ a_q, \ g_q,\ \theta \right].
\end{equation}
Here $r_q = \rex / \rin$. The eccentricity $\epsilon$ is defined as $\epsilon = d/\left(\rex - \rin \right)$. The fading parameter $f$ is defined as $f =  h_0/h$. It specifies ``non-contrastness" (or fading) of the ring image: when $f = 0$ the brightness of the eccentric ring grows from zero to unity, while when $f = 1$, the brightness is uniform. The additional elliptical Gaussian brightness is specified with three parameters: $\gax$, $a_q$, and $g_q$. The main axis of the FWHM ellipse, $\gax$, is expressed in $\rex$, i.e. $\gax = a/\rex$. The main axis is perpendicular to the $X$ axis when $\theta = 0$. The ellipse axial ratio is $a_q = b/a$. The fraction of the Gaussian flux in the total flux is controlled by the $g_q$ parameter: $g_q = 0$ corresponds to a model without an additional Gaussian component, while $g_q = 1$ means that $\zsp$ is entirely due to the Gaussian component. 

The zero-spacing flux of the model (without the Gaussian) is equal to the integral over the area containing the slashed eccentric ring, or the volume of the geometric figure shown in Fig.~\ref{nine_param_geometry}b. By analogy with Eqn.~\ref{unity_zsp_h}, we find the maximal brightness $h$ that brings the volume to unity as
\begin{eqnarray}
  \label{unity_zsp_h_9pm}
  h = \frac{2}{\pi} \Biggl[ \left(\rex^2 - \rin^2 \left(1 + \frac{d}{\rex} \right) \right) f + \nonumber \\ 
     \left(\rex^2 - \rin^2 \left(1 - \frac{d}{\rex}  \right) \right) \Biggr]^{-1}.
\end{eqnarray}
The eccentric ring function can be defined as
\begin{equation}
  \label{xring_bri}
  \xringf \left(r \right) = 
  	  \circf \left( \frac{r}{\rex} \right) - \circf\left( \frac{r-d}{\rin} \right),
\end{equation}
or, in the visibility domain,
\begin{eqnarray}
  \label{xring_vis}
  \xringf\left(\rho \right) &=& \frac{\rex J_1(2 \pi \rex \rho)}{\rho} \nonumber \\
  		&-&  e^{i 2 \pi d u}\frac{\rin J_1(2 \pi \rin \rho)}{\rho}.
\end{eqnarray}
The pillboxes are slashed down to the $h_0$ brightness, so it may be non-zero at the darkest side of the ring. This slashing is fulfilled by multiplying Eqn. \ref{xring_bri} by the linear function
\begin{equation}
  \label{lin_fun_h0}
  l(x) = \frac{1}{2} \left( \frac{h - h_0}{\rex} x + h + h_0 \right).
\end{equation}
This results in a brightness change similar to that given by Eqn.~\ref{prod_lxring}, but in this model the maximum brightness is $h$ at the point $(\rex,0)$ and the minimum brightness is $h_0$ at $(-\rex,0)$. 
For brevity we introduce variables $\xi = (h+h_0)/2$ and $\eta = (h-h0)/2$. Using the same notation as Eqn.~\ref{ft_of_circ} in the previous section for the FTs of the pillboxes, $F(\rho,\rex) \rightleftharpoons \circf(r/\rex)$ and $F(\rho,\rin) \rightleftharpoons \circf(r/\rin)$, and the formula Eqn.~\ref{F_der} for their derivatives, we can write expressions for the visibilities of external and internal slashed pillboxes as
\begin{equation}
  \label{vis_ex_9prm}
    \vex(u,v) = \xi F(\rho,\rex) + \frac{i}{2 \pi} \eta \frac{\ud }{\ud \rho} F(\rho,\rex)u
\end{equation}
and
\begin{eqnarray}
  \label{vis_in_9prm}
    \vin(u,v) &=& \left(\xi - \eta\frac{d}{\rex} \right) F(\rho,\rin)    \nonumber \\
              &+& \frac{i}{2\pi} \eta \frac{\rin}{\rex} \frac{\ud }{\ud \rho} F(\rho,\rin)u.
\end{eqnarray}
The visibility of the slashed eccentric ring is thus expressed as the difference
\begin{equation}
  \label{slxring_vis}
  V_\mathrm{r}(u,v) = \vex(u,v) - e^{i 2\pi d u} \vin(u,v).
\end{equation}
The Gaussian is centered at the inner edge of inner ring, at $x = \rin - d$, so the shift factor is $\exp(i 2\pi u (\rin - d))$, and the visibility of the Gaussian according to Eqn. \ref{gaussian_transform} is
\begin{equation}
  \label{gauss}
  V_\mathrm{g}(u,v) = e^{-2 \pi^2 k^2 \left((u a)^2 + (v b)^2 \right) - i 2\pi u (\rin - d)}.
\end{equation}
where $k$ is the coefficient transforming FWHMs $a$ and $b$ into the standard deviations $\sigma_a$ and $\sigma_b$ according to Eqn.~\ref{fwhm}:
\begin{equation}
  \label{gauss}
  k = \frac{1}{2\sqrt{2 \ln 2}}
\end{equation}
Both the Gaussians defined in Eqns.~\ref{gaussian_transform} and \ref{gauss} and the slashed eccentric ring in Eqn.~\ref{slxring_vis} have unity integrals over the $xy$-plane. Therefore, they are easily combined to create the nine-parameter model visibility as
\begin{equation}
  \label{final_9param}
  V(u,v) = \zsp\left(\left[1-g_q\right]V_\mathrm{r} + g_q V_\mathrm{g} \right),
\end{equation}
where $g_q \in [0,1]$ is the fraction of the total flux contained in the Gaussian component.

The model image
orientation is determined by the rotation angle $\theta$ used in the coordinate transformation from $(u,v)$ to $(u',v')$ given in Eqn.~\ref{rotate_uv}. 

\begin{figure}[ht!]
\plotone{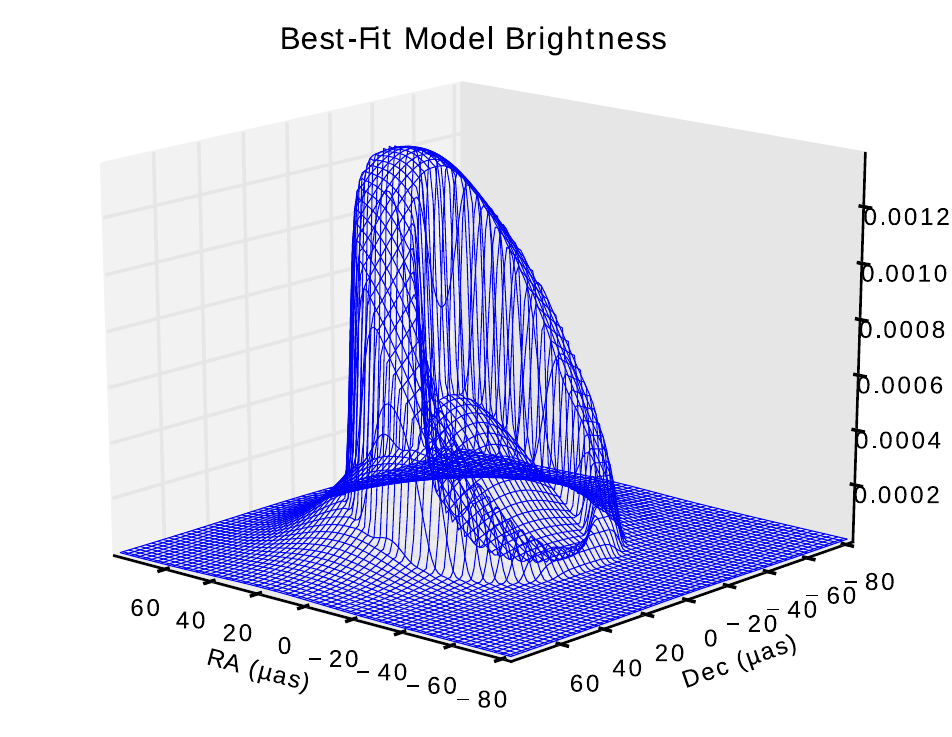}
\caption{\small A three-dimensional wire-frame image of the nine-parameter model brightness function. The Gaussian component causes the brightness to fall off more smoothly beyond the slashed ring.\label{f4_bestfit_wframe}}
\end{figure}

The nine-parameter model brightness is visualized in Fig.~\ref{f4_bestfit_wframe}.

\section{Markov Chain Monte-Carlo with Replica Exchange}

\subsection{Bayesian Inference}

\begin{figure*}
  \begin{center}
        \pdfimageresolution=350
    \includegraphics{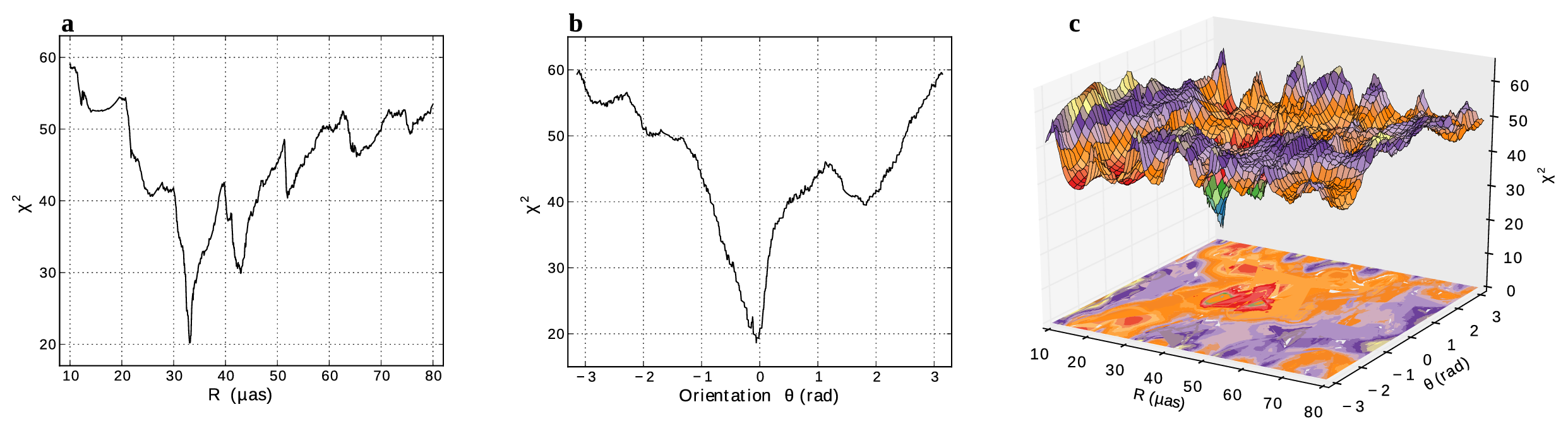}
  \end{center}
\caption{\small Panels ${\bf a,b}$: one-dimensional cross-sections of the $\chi^2$ surface with one parameter varying and the other eight parameters being constant near the global minimum. ${\bf a}$: dependence of $\chi^2$ on the external radius, $R_{\rm ex}$, of the model crescent, $\chi^2(R_{\rm ex}) = \chi^2(2.33, R_{\rm ex}, 0.92, 0.6, 0.002, 1.37, 0.82, 0.58, 0.)$. ${\bf b}$:  dependence of $\chi^2$ on the orientation angle $\theta$ of the model crescent: $\chi^2(\theta) = \chi^2(2.33, 33.0, 0.92, 0.6, 0.002, 1.37, 0.82, 0.58, \theta)$. Panel ${\bf c}$: two-dimensional cross-section of the $\chi^2$ surface with two parameters, $R_{\rm ex}$ and $\theta$, varying and the other seven parameters frozen near the global minimum: $\chi^2(R_{\rm ex},\theta) = \chi^2(2.33, R_{\rm ex}, 0.92, 0.6, 0.002, 1.37, 0.82, 0.58, \theta)$. The complexity of $\chi^2$ and large number of local minimums make the gradient methods of global optimization unfeasible.
\label{chi2_rad_ori_3d}}
\end{figure*}

The problem of finding a tuple of model parameters
(e.g., $\bu = \left[ \zsp, \rex, \rin, \ d, \ f, \ a, \ b, \ g_q, \ \theta \right]$ for the nine-component model)
that provide the best approximation to the observation data is an optimization problem that cannot be solved with the use of gradient methods. They were developed for smooth functions with a small number of local minima. Here we need to find the global minimum of the $\chi^2$ distribution
\begin{equation}
  \label{chi2_distr}
  \chi^2 = \chi_{\rm vis}^2 + \chi_{\rm clp}^2 ,
\end{equation}
where $\chi_{\rm vis}^2$ determines the deviation of the observed $N_{\rm vis}$ visibility amplitudes from that of the model,
\begin{equation}
  \label{chi2_vis_distr}
  \chi^2_{\rm vis}= \sum_{i=1}^{N_{\rm vis}} \frac{(|V_i^{\rm obs}| - |V_i^{\rm mod}|)^2}{\sigma_v^2},
\end{equation}
and $\chi^2_{\rm clp}$ is the same but for the $N_{\rm clp}$ closure phases:
\begin{equation}
  \label{chi2_clp_distr}
  \chi^2_{\rm clp}= \sum_{i=1}^{N_{\rm clp}} \frac{(\Psi_i^{\rm obs} - \Psi_i^{\rm mod})^2}{\sigma_\Psi^2},
\end{equation}
Here $\sigma_v^2$ and $\sigma_\Psi^2$ are the respective standard deviations.
The $\chi^2(\zsp, \rex, \rin, \ d, \ f, \ a, \ b, \ g_q, \ \theta)$ thus defined is a nine-dimensional hypersurface with a tremendous number of local ``creases". Fig~\ref{chi2_rad_ori_3d} gives an example of one- and two-dimensional $\chi^2(\zsp, \rex, \rin, \ d, \ f, \ a, \ b, \ g_q, \ \theta)$ cross-sections along the $R_{\rm ex}$ and $\theta$ model parameters and over the $(R_{\rm ex},\theta)$ plane. A gradient method will most probably stop at a local minimum quite far from the global one.  On the other hand, an exhaustive search over the nine-dimensional grid is too computationally intensive.

That said, statistical methods could be more helpful for the model parameter estimation. We use one of the most powerful statistical methods, the Bayesian inference. In the Bayes paradigm, the new information, the ``evidence", is used to update the ``prior" guess on the probability of a hypothesis, with the use of the well known Bayes' theorem
\begin{equation}
  \label{bayes_rule}
  P(A|B) =  \frac{P(B|A)P(A)}{P(B)}.
\end{equation}
From the prior probability of event $A$, $P(A)$, its posterior probability, $P(A|B)$, is inferred, having the evidence $P(B)$ and the likelihood $P(B|A)$. Instead of single events $A$ and $B$ and their (conditional) probabilities the probability distributions defined on generally multidimensional spaces can be considered. Bayes' theorem (Eqn.~\eqref{bayes_rule}) can be used to calculate the posterior distribution at any point using the values of the three other distributions. This is used to obtain the posterior distribution of the fitted model parameters from the distribution of the observation data and the ``prior" model parameter distribution. Numerically it can be implemented as  probing the parameter space evenly enough to get the sufficiently dense set of points to plot the histogram of posterior distribution. Its maximum (or maxima) will be at the best-fit parameter values. 

We assume that the specific set of observed data $\bx$ (visibility amplitudes and closure phases) is a sample from the multidimensional random variable, ${\bf X}$, with the probability density distribution $P({\bf X})$. For a sample $\bx$, $P(\bx)$ is a single number, the value of $P({\bf X})$ at the point $\bx$. Further lower case letters are used instead of the capitals, so $P(\bx)$ actually means $P({\bf X})$.

Within the Bayesian framework both the observed data set $\bf{x}$ and the model parameter tuple $\bu$ are considered as statistically linked multi-dimensional random variables with their joint probability distribution
\begin{equation}
  \label{joint_ux}
  P(\bu,\bx) =  P(\bu|\bx)P(\bx) = P(\bx|\bu)P(\bu),
\end{equation}
where ``," reads ``and". Relationship \eqref{joint_ux} associates the probability densities named as follows:
\begin{itemize}
  \item[-] $P(\bx)$ the evidence,
  \item[-] $P(\bu)$ the prior or prior distribution,
  \item[-] $P(\bx|\bu)$ the likelihood, and
  \item[-] $P(\bu|\bx)$ the inference or the posterior probability distribution.
\end{itemize}
In terms of causality, the object under observation is the cause, and the observation data is the effect. The Bayes' theorem allows us to rearrange the cause and the effect: using the known data $\bx$, compute the \emph{posterior} probability distribution $P(\bu|\bx)$ that $\bx$ is an effect of the object represented by our model with the parameter set $\bu$. Thus, we pose a task to find not just a single set of the ``optimal" model parameters, but the probability distribution of this set over the parameter space \emph{given} the actual set of observation data. Of course, we are interested in such distributions for every single parameter, which are the marginal distributions of $P(\bu|\bx)$. Below is shown that MCMC allows direct rendering of these marginal distributions. The likelihood $P(\bx|\bu)$ may be any positive function that reaches its maximum when the difference between the actual data and the model data becomes zero. We use a Gaussian likelihood
\begin{equation}
  \label{gaussian_likelihood}
  P(\bx|\bu) = \left(\prod_i \frac{1}{\sqrt{2 \pi} \sigma_i} \right) e^{-\frac{1}{2}\chi^2},
\end{equation}
where $\chi^2$ and $\sigma$ are from Eqs.~\eqref{chi2_distr}--\eqref{chi2_clp_distr}.
The prior, $P(\bu)$, is the distribution over the parameter space that represents our preliminary knowledge about the intervals where the parameter values could be present.
The prior may not be very informative (for example, a uniform value within the allowed interval and zero outside), but it must always be provided.

Dividing \eqref{joint_ux} by $P(\bx)$ yields the Bayes' formula:
\begin{equation}
  \label{bayes_formula}
  P(\bu|\bx) =  \frac{P(\bx|\bu)P(\bu)}{P(\bx)},
\end{equation}
with the searched for posterior parameter distribution on the left hand side, and computable probabilities on the right hand side. The value of $P(\bx)$, the probability density of the given observation data sample, can be calculated using the total probability law,
\begin{equation}
  P(\bx) = \int P(\bx|\bu)P(\bu) \ud \bu.
\end{equation}
For a given prior and a model the evidence $P(\bx)$ is always a single constant value as long as we work with the same data set: the integration over the whole parameter space removes all the variables. The evidence value can be used to compare the quality of different models. A ``better" model will have larger $P(\bx)$. The Bayes' theorem thus takes the form
\begin{equation}
  \label{post_distr_int}
  P(\bu|\bx) =  \frac{P(\bx|\bu)P(\bu)}{\int P(\bx|\bu)P(\bu) \ud \bu}.
\end{equation}
As we already said, the posterior distribution of all the parameters, $P(\bu|\bx)$, is not as interesting as that of an individual parameter, $P(u_i|\bx)$, $u_i \in \bu$, because it can provide the information on the mean value (or values, if multi-modal) and uncertainty of the estimate of the parameter $u_i$. Such individual distributions for every parameter $u_i$ are, in effect, the marginal distributions, i.e. the results of integration of the total distribution $P(\bu|\bx)$ over the parameter subspace spanned by all the parameters but $u_i$:
\begin{equation}
  \label{param_post_distr}
  P(u_i|\bx) =  \int P(\bu|\bx)\ud u_1 \ud u_2 ... \ud u_{i-1} \ud u_{i+1} ... \ud u_N.
\end{equation}
The posterior distributions, $P(u_i|\bx)$, are not required to be normalized, so the strict equations~\eqref{bayes_formula} or~\eqref{post_distr_int} can be relaxed to a mere proportionality
\begin{equation}
  \label{bayes_propor}
  F(\bu|\bx) \propto P(\bx|\bu)P(\bu),
\end{equation}
where $F(\bu|\bx) \propto P(\bu|\bx)$. Normalization of the $F$ function would produce the posterior distribution $P(\bu|\bx)$ and its marginals $P(u_i|\bx)$. However, the statistical parameters of $P(u_i|\bx)$---means and standard deviations, or qualitative conclusions about their forms---can be found directly from $F(u_i|\bx)$ without the normalization. The Metropolis-Hastings algorithm described here utilizes this fact. It draws many samples from the $P(u_i|\bx)$ distributions, and the result of optimization, $\bu$, is obtained from the histograms built using the saved samples. 

\subsection{Metropolis-Hastings Algorithm}

In order to apply the Bayesian inference method to the problem of finding the best-fit model parameters we use a strong algorithm named Markov Chain Monte Carlo (MCMC) with Replica Exchange (or Parallel Tempering).  The algorithm has three stages. First, an initial set of parameters $\bu_0$ is randomly drawn from the prior distribution $P(\bu)$. The two other stages, the burn-in and the search, are essentially the same except at the burn-in stage the optimal steps for each parameter are picked. The iterations generate the Markov chain of the parameter tuples $\bu_i = \left(p_{i,1}, p_{i,2}, ... p_{i,N_p},  \right)$, and the more iterations, the better the $\bu_i$ values approximate $P(\bu|\bx)$. The Markov property, i.e. the dependence of the $i_{th}$ chain element on the previous $(i-1)^{th}$ element only is ensured by the method of their generation. At each iteration, a \emph{proposal} model parameter set $\bp^\prime$ is generated from the proposal distribution $q(\bp_{i-1};\bp^\prime)$. The new proposal set is randomly accepted or rejected with a probability $\alpha$,
\begin{equation}
  \label{accept_probab}
  \alpha =  \min \left( \frac{P(\bx|\bu')P(\bu')q(\bu_{i-1};\bu')} 
            {P(\bx|\bu_{i-1})P(\bu_{i-1})q(\bu';\bu_{i-1})}, 1 \right).
\end{equation}
In the Metropolis-Hastings algorithm the proposal distribution $q(\bu_i;\bu_j)$ \emph{must} be symmetric. Here it is assumed a Gaussian distribution
\begin{equation}
  \label{gaussian_prop_distr}
  q(\bu_i;\bu_j) = \prod^{N_p}_{k} \frac{1}{\sqrt{2 \pi \sigma^2_k}} 
                   \exp \left( \frac{(p_{j,k} - p_{i,k})^2}{2 \sigma^2_k} \right).
\end{equation}
Since $q(\bu_{i-1};\bu') \equiv q(\bu';\bu_{i-1})$, the acceptance probability is simplified to
\begin{equation}
  \label{accept_probab2}
  \alpha =  \min \left( \frac{P(\bx|\bu')P(\bu')} {P(\bx|\bu_{i-1})P(\bu_{i-1})}, 1 \right).
\end{equation}
Obviously, the numerator and denominator in~\eqref{accept_probab2} are the right hand sides of~\eqref{bayes_propor} for the new and previous $\bp$, respectively, which in turn are proportional to the desired probability distribution. If the probability of proposal $\bp^\prime$ set is greater, then $\alpha = 1$, and $\bu^\prime$ becomes the new parameter set unconditionally. Due to the Gaussian likelihood, i.e. uncertainty in the observations~\eqref{gaussian_likelihood}, theacceptance probability $\alpha$ becomes
\begin{eqnarray}
  \label{accept_probab3}
  \alpha =  \min \biggl( \exp \left\lbrace -\frac{1}{2}\left( \chi^2(\bx;\bu') - 
                        \chi^2(\bx;\bu_i) \right)  \right\rbrace     \times              \nonumber \\
            \frac{P(\bu')} {P(\bu_i)}, 1 \biggr) .    
\end{eqnarray}
If the proposed parameter set were accepted only in case $\alpha=1$ when the new point is necessarily better (with lower $\chi^2$) than the previous one, the algorithm would be the basic random Monte-Carlo search. Unfortunately, the basic random search suffers from the ``curse of dimensionality": the rejection probability exponentially grows with the number of dimensions. Hence, a basic random search of many parameters will last forever. \citet{Metropolis_etal_1953} suggested a way out: accept not only $\chi^2$-better parameter sets, but also the sets that worsen $\chi^2$, but accept it with the probability $\alpha$. This technique ensures the ``random walk" of $\bu_i$, exploring the parameter space and visiting the volumes with better posterior probability more frequently than others. If the proposal set is rejected, the previous state will be repeated in the chain.

For models with many parameters the acceptance probability $\alpha$ tends to become small if all the parameters are stepped simultaneously, lowering the rate of acceptance and the overall algorithm efficiency. For this reason at each iteration we step only one parameter, keeping others constant. The following pseudocode describes one MCMC algorithm iteration:
\begin{enumerate}
\item Randomly choose $u_{i-1,j}$ from $\bu_{i-1}$, parameter number $j$ uniformly distributed;
\item Generate the $j$-th proposal parameter $u_{i-1,j}$ from the Gaussian distribution;
\item Calculate $\alpha$ and accept or reject $u_{i-1,j}$ with probability $\alpha$;
\item Repeat 1-3 for $N_p$ times; Memorize the newly generated state as $\bu_i$.
\end{enumerate}

The efficiency of this algorithm is also sensitive to the step size of proposal distribution~\eqref{gaussian_prop_distr}, which is determined by the variance of the Gaussian distribution. If it too small, most of the trial points are accepted, but the random walk is too slow to sample all the parameter space. Conversely, if the step is too large, most of the trial points are rejected and the MCMC algorithm can get stuck at a certain point for a long time despite the ability to make large jumps. Previous empirical studies recommend optimizing the step size to make the accept rate $\sim 25$\% in high-dimensional cases \citep[see references in][]{Gregory2005}. The second, burn-in stage of MCMC is intended to adaptively adjust steps for all the parameters. After updating a $j^{th}$ parameter $u_{i,j}$, if the accept rate of newest 100 trials is more than 30\%, then the variance $\sigma_j$ is multiplied by 1.01. Otherwise, if the accept rate of newest 100 trials is less than 20\%, the variance $\sigma_j$ is divided by 1.01.

\subsection{Replica Exchange MCMC Algorithm}

The described Metropolis-Hastings MCMC algorithm is quite suitable for our problems where the direct sampling is complicated or impossible. However, a simple Metropolis-Hastings MCMC algorithm can fail to fully explore the target probability distribution, especially if the distribution is multi-modal with widely separated peaks. The algorithm can get trapped in a local mode and miss other regions of parameter space that contain significant probability. 

The replica-exchange MCMC algorithm (also known as parallel tempering) is a result of improvement of the MCMC algorithm targeted to such complex multi-modal distributions. The replica-exchange algorithm belongs to the class of ``generalized-ensemble algorithms". It has been developed mostly in the past decade and recently was applied to some astronomical problems \citep{Gregory2005,Varghese_etal2011,Benneke_Seager2012}. In this method a parameter $\beta$ called ``temperature" is introduced as
\begin{eqnarray}
  \label{post_distr_beta}
  P(\bu|\bx;\beta) &=&  \frac{P(\bx|\bu)^\beta P(\bu)}{\int_{\bu} \int_{\beta} 
                              P(\bx|\bu)^\beta P(\bu) \ud \bu \ud \beta} \nonumber \\
      &\propto& P(\bx|\bu)^\beta P(\bu)  .
\end{eqnarray}
When $\beta=1$, it becomes the target posterior distribution. For the Gaussian likelihood \eqref{gaussian_likelihood} the latter can be rendered as
\begin{equation}
      P(\bu|\bx;\beta) \propto \exp \left( \beta L(\bx|\bu) \right) P(\bu),
\end{equation}
where $L(\bx|\bu)$ is a log-likelihood. The term ``temperature" is borrowed from the canonical distribution $\exp \left( -\beta E \right)$ in statistical mechanics, where the absolute temperature is expressed using the ``thermodynamic $\beta$" written as 
\begin{equation}
	\beta = \frac{1}{kT},
\end{equation}
so $\beta$ is inversely proportional to the temperature. Using this analogy one can see that in Eq.~\eqref{post_distr_beta}  the log-likelihood $L(\bx|\bu)$ plays the role of negative energy $-E$. High temperature (means low $\beta$) makes the likelihood function flatter and also makes the Metropolis-Hastings acceptance probability $\alpha$ higher, because
\begin{eqnarray}
  \label{accept_probab_beta}
  \alpha = \min \biggl( \exp \left\lbrace -\frac{\beta}{2}\left( \chi^2(\bx;\bu') - \chi^2(\bx;\bu_i) \right)
              \right\rbrace \nonumber \\
         \times  \frac{P(\bu')} {P(\bu_i)}, 1 \biggr).  
\end{eqnarray}
Thus, the Metropolis-Hastings sampling at higher temperatures enables exploration of wider ranges of the parameter space. 

In the replica exchange MCMC algorithm, multiple Markov chains with different temperatures $(\beta_1,\beta_2, ... \beta_{N_\beta})$ including a chain with the lowest temperature $\beta=1$ and different initial conditions are generated in parallel. The specific values of $\beta_l$ usually span several orders of magnitude with logarithmic steps. As an example, 40 Markov chains may have $\beta_l \in [10^{-4} ; 1]$. At each MCMC iteration, when the generation of new sets of parameters $\bu_{i,l}$ is finished in all the chains, the newly generated elements of adjacent chains at the temperatures $\beta_l$ and $\beta_l$ are exchanged with a probability $\alpha$ written as
\begin{equation}
  \label{exchange_probab}
  \alpha =  \min \left( \frac{P(\bu_{i,l+1}|\bx; \beta_{l+1})} 
            {P(\bu_{i,l}|\bx; \beta_{l})}, 1 \right) .
\end{equation}
The exchange procedure is repeated for $N_\beta - 1$ times, after which a new parameter $\bu_{i+1}$ generation begins. Under the Gaussian  likelihood \eqref{gaussian_likelihood} and the Gaussian proposal distribution \eqref{gaussian_prop_distr} it becomes
\begin{eqnarray}
  \label{accept_probab_beta2}
  \alpha &=& \min \biggl( \exp \biggl\lbrace -\frac{1}{2}(\beta_{l+1} - \beta_{l}) \nonumber \\
         &&\times  \Bigl( \chi^2(\bx;\bu_{i,l}) -
                 \chi^2(\bx;\bu_{i,l+1}) \Bigr) \biggr\rbrace \frac{P(\bu')}
                       {P(\bu_i)}, 1 \biggr)      .
\end{eqnarray}
In the higher temperature distributions $(\beta \ll 1)$, radically new configurations are explored, while lower temperature distributions  $(\beta \approx 1)$ allow for detailed exploration of new configurations and local modes. The final inference on the model parameters is based on samples drawn from the target probability distribution $(\beta=1)$ only.


\section{Simulation Setup}

\begin{figure}[ht!]
\epsscale{0.7}
\plotone{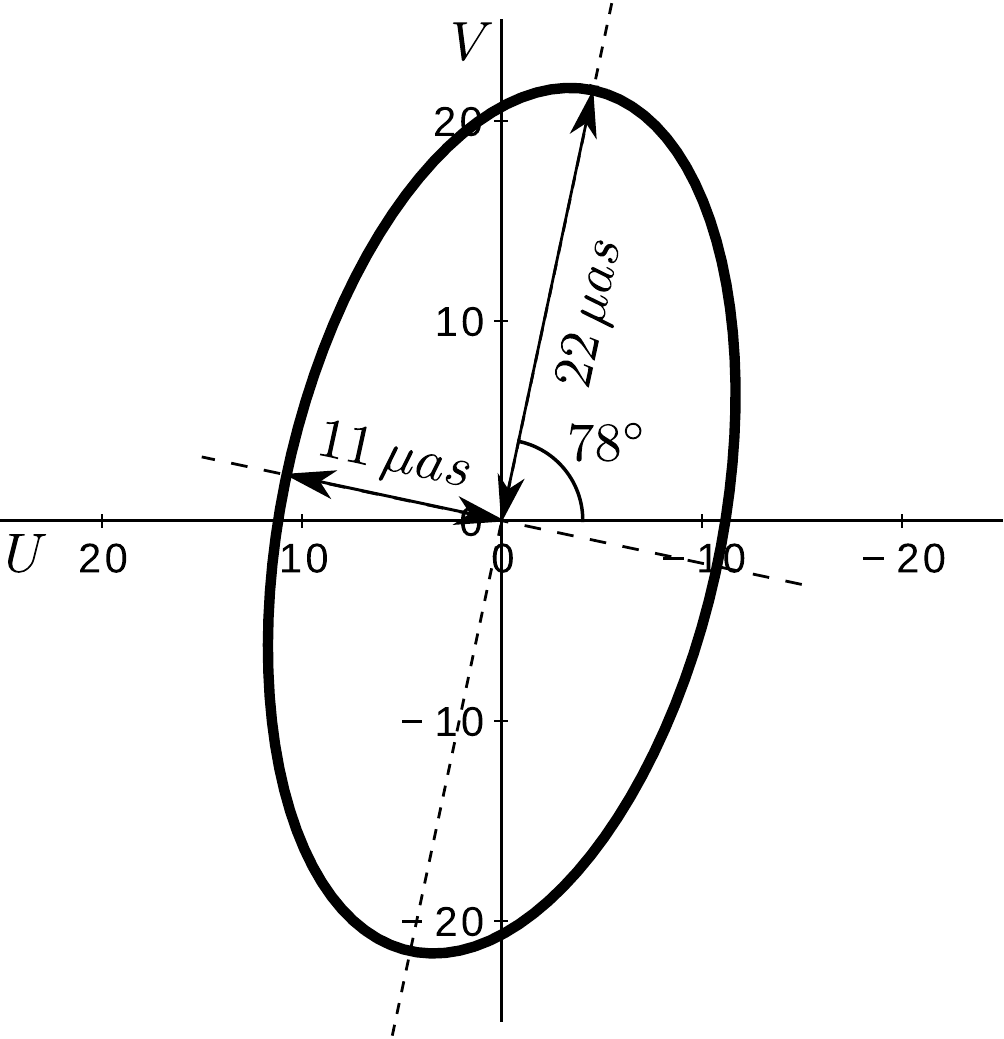}
\caption{\small The convolution kernel: shown the elliptical FWHM locus of the two-dimensional Gaussian kernel used for artificial scattering of the Sgr A* images.
\label{fig_scattering_ellipse}}
\end{figure}

To test the usability of the nine-parameter model for imaging of the Sgr A* black hole in different possible states, we conducted a series of simulated EHT observations of the Sgr A* images using the MAPS package. We used for observations the set of Sgr A* images simulated with the use of the BJPL2013 physical model developed by A.~E. Broderick, T. Johannsen, D. Psaltis, and A. Loeb \citep{BJPL2013}. In order to imitate the scattering by the turbulent ionized interstellar medium, the images were smoothed by convolving with an elliptical Gaussian kernel with a FWHM of 22~$\mu$as along the major axis and 11~$\mu$as along the minor axis, with a position angle of 78$^\circ$ \citep[see][]{Bower2006,Shen2005,Bower_etal2004,Falcke2000}. The elliptical locus of the 2D Gaussian FWHM is sketched in Fig.~\ref{fig_scattering_ellipse}.

In the simulated observations a VLBI array of the following eight radio telescopes was used: Manua Kea, Hawaii; SMTO (Sub-Millimeter Telescope Observatory), Arizona; CARMA (Combined Array for Research in Millimeter-wAve), California; LMT (Large Millimeter Telescope), Mexico; ALMA (Atacama Large Millimeter/submillimeter Array), Chile; Pico Veleta, Spain; Plateau de Bure, France; and SPT (South Pole Telescope). The antenna parameters used in simulation are provided in Tab~\ref{tab_EHT}.

\begin{deluxetable*}{lrrrrrcccr}[ht!]   
  \tablecaption{Antennae Comprising the Event Horizon Telescope \label{tab_EHT}}
  \tablewidth{0pt}
  \tablehead{ 
    & \colhead{Geocentric} & \colhead{Geocentric} & \colhead{Geocentric} & \colhead{Lat.} & \colhead{Lon.} & &
         \colhead{Low} & \colhead{High} & \colhead{SEFD} \\ 
    \colhead{Name} &  &  &  &  &  & \colhead{Dish (m)} & \colhead{Elevation} & \colhead{Elevation} &  \\
    \vspace{0.1cm}
  & \colhead{$X$ (m)} & \colhead{$Y$ (m)} & \colhead{$Z$ (m)} & \colhead{($^\circ$)} & 
         \colhead{($^\circ$)} &  & \colhead{($^\circ$)} & \colhead{($^\circ$)} & \colhead{(Jy)}
    }
  \startdata
  Hawaii8 & -5,464,523.4000 & -2,493,147.0800 &  2,150,611.7500 &  19.8244 & -155.4734 & 20.8 & 15 & 85 & 3,300  \\
  SMTO    & -1,828,796.2000 & -5,054,406.8000 &  3,427,865.2000 &  32.7016 & -109.8912 &  10  & 15 & 85 & 11,900 \\
  CARMA8  & -2,397,431.3000 & -4,482,018.9000 &  3,843,524.5000 &  37.2314 & -118.2892 & 26.9 & 15 & 85 & 7,500  \\
  LMT     &   -768,713.9637 & -5,988,541.7982 &  2,063,275.9472 &  18.9859 &  -97.3149 &  50  & 15 & 85 & 4,000  \\      
  ALMA50  &  2,225,037.1851 & -5,441,199.1620 & -2,479,303.4629 & -23.0279 &  -67.7549 & 84.7 & 15 & 85 &   110  \\
  PV      &  5,088,967.9000 &   -301,681.6000 &  3,825,015.8000 &  37.0662 &   -3.3926 &  30  & 15 & 85 & 2,900  \\
  PdBI    &  4,523,998.4000 &    468,045.2400 &  4,460,309.7600 &  44.6339 &    5.9067 & 36.7 & 10 & 85 & 1,600  \\         
  SPT     &          0.0000 &          0.0000 & -6,359,587.3000 & -90.0000 &    0.0000 &  12  & 15 & 85 & 10,000 
  \enddata
\end{deluxetable*}

\begin{figure}[ht!]
  \begin{center}
        \pdfimageresolution=600
    \includegraphics{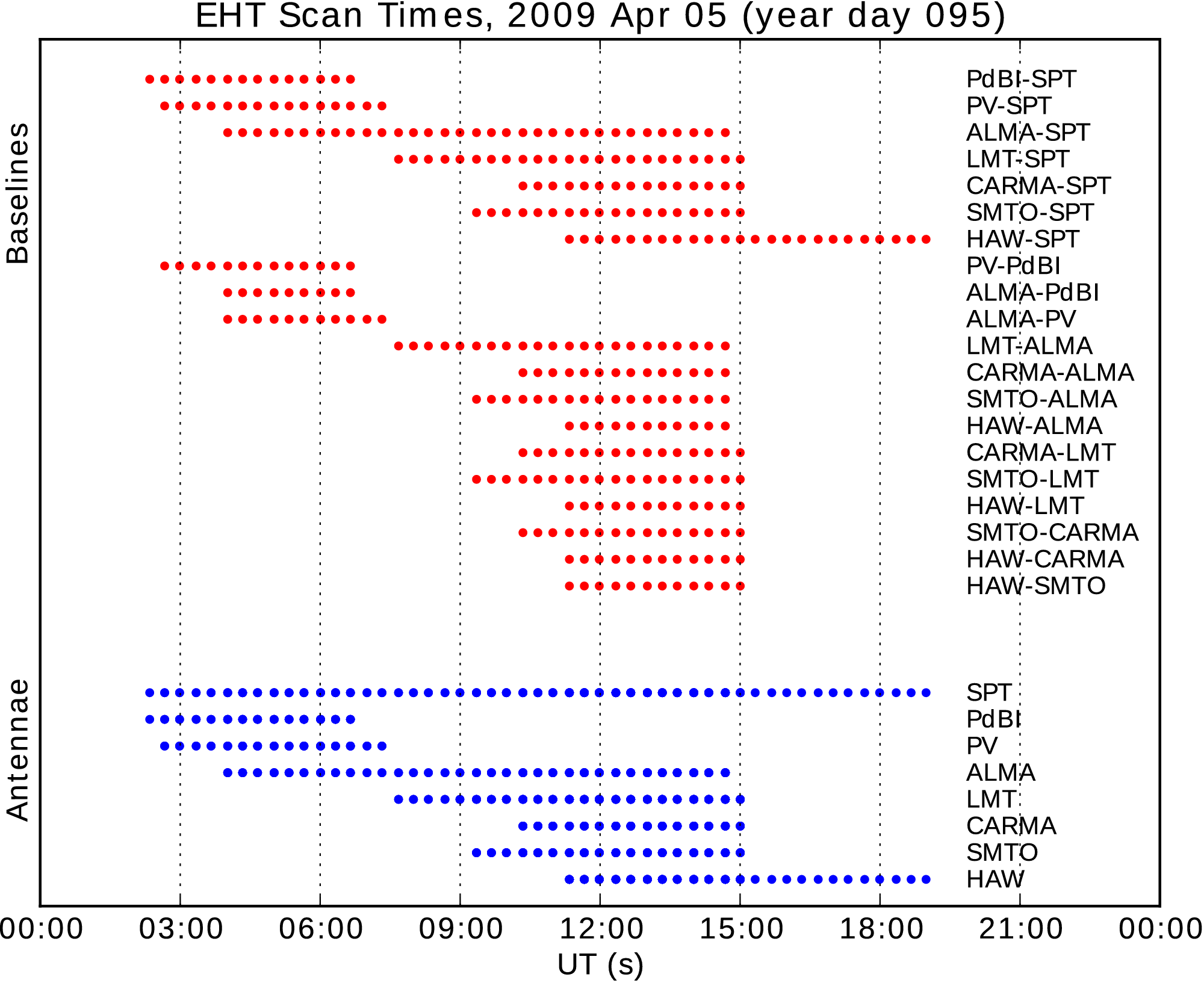}
  \end{center}
\caption{\small Availability of the EHT antennas and baselines over the full track of the simulated Sgr A* observations. The scans follow each 20 minutes. Each scan lasts 1 minute, integration time is 1 minute. 
\label{fig_eht_scan_times}}
\end{figure}

The MAPS software package (MIT Array Performance Simulator) was originally developed at the MIT Haystack observatory. It is a versatile tool used for simulating work of any interferometer. In particular, for a given brightness distribution, a radio telescope array structure, frequency channels, scan durations and integration times, MAPS creates a full set of visibilities both in ASCII and in the standard UVFITS format. Tab~\ref{tab_EHT} is \emph{per se} one of the MAPS input files. An early example of using MAPS is in \citet{ska2004hall}. \citet{RusenLu2014} used MAPS to obtain visibility data for model-independent imaging of Sgr A* and M87 galactic centers. The observations were simulated for a full track (24 hours), with 1-minute scans repeating every 20 minutes with 1-minute integration times at the frequency 229.089 GHz. The correlator channel bandwidth 500 MHz.  Thermal noise for these parameters was included. Sgr A* is only visible from a subset of the array at any given time. The availability of the antennas and the baselines over the full track is shown in Fig.~\ref{fig_eht_scan_times}. This provided fairly good $uv$-coverage with the total of 329 visibility and 325 closure phase values.

In this simulation we have used images with a small number of pixels, only 100$\times$100 for the BJPL2013 model with the pixel size 100 $\mu$as. In the $uv$ plane this implies the huge respective pixel sizes of 2 G$\lambda$ and 0.8 G$\lambda$. Each scan samples the visibility over the patches of $uv$ plane, whose sizes are determined by the bandwidth (0.5 GHz) and integration time (1 minute). With the short wavelength ($\lambda=1.3095$ mm) and very long baseline lengths (up to $9\times 10^{9} \lambda$) the patch sizes vary from ~1.5 M$\lambda$ to ~15 M$\lambda$, growing with the baseline length. Typically, the patches are 100-1000 times smaller than the pixels, and a single pixel can contain multiple patches. The oversampling by zero-padding the observed brightness image to increase its grid to, say, 2048 nodes can make the $UV$-plane pixels finer: 100 M$\lambda$. However, even the 8192$\times$8192 grid reduces the pixel size to only 25 M$\lambda$, which is still almost twice as large as the largest patch. Also, large grids exponentially increase the computation time. Fortunately, the grid size appears to exert negligible influence on the simulation results because MAPS never samples the visibility value of a single pixel. Instead, it makes 2D spline interpolation over the 3$\times$3 pixel vicinity of every pixel under the patch, thus providing effective ``scalability" of the grids.   

Before trying to reconstruct the image from these data points we had an option to ``descatter" them. The descattering is performed by multiplying the observed visibilities at $UV$-points by the \emph{inverse} of the scattering kernel, which is equivalent to their deconvolution in the brightness domain.  


The method has been tested on several Sgr A* images computed on two different physical models. The images provided by \citet{BJPL2013} are based on the BJPL2013 model developed by A.~E. Broderick, T. Johannsen, D. Psaltis, and A. Loeb. A second set of images was received from J. Dolence and M. Moscibrodska \citep[see][]{Monica2009radmod,Monica2009magnaccr,Monica2011numeric,Dolence2012numosc,Shiokawa2012relmhd}. 

The BJPL2013 model images can be ordered by three parameters: black hole spin, $a$ (M), its inclination angle, $i\; (^\circ)$, and $\epsilon$, the residual (non-GR) quadrupole moment. It is a parameter of the suggested non-Kerr space-time metric $Q = -M(a^2 + \epsilon M^2)$ and it is the measure of deviation from the General Relativity (GR). When $\epsilon = 0$, the space-time metric is the Kerr metric, and the no-hair theorem is true. It has been shown \citep{deVries2000,Vries2005limacon,Johannsen2010testbhim,Johannsen2010testqk,Johannsen2012testsgra} that non-zero $\epsilon$ deforms the shadow, making it deviate from a circle at $\epsilon = 0$. Our model allows only circular shadows, so we do not use BJPL2013 images with non-zero $\epsilon$.


\section{Simulation Results}

We aim at assessing the similarity between the simulated images and the 9-parameter model images and estimate the shadow size and the spin. For the Sgr A* images the comprehensive physical model BJPL2013 was used.

\subsection{Observations of 9-Parameter Model Itself} 

\begin{figure*}[ht!]
  \begin{center}
        \pdfimageresolution=350
    \includegraphics{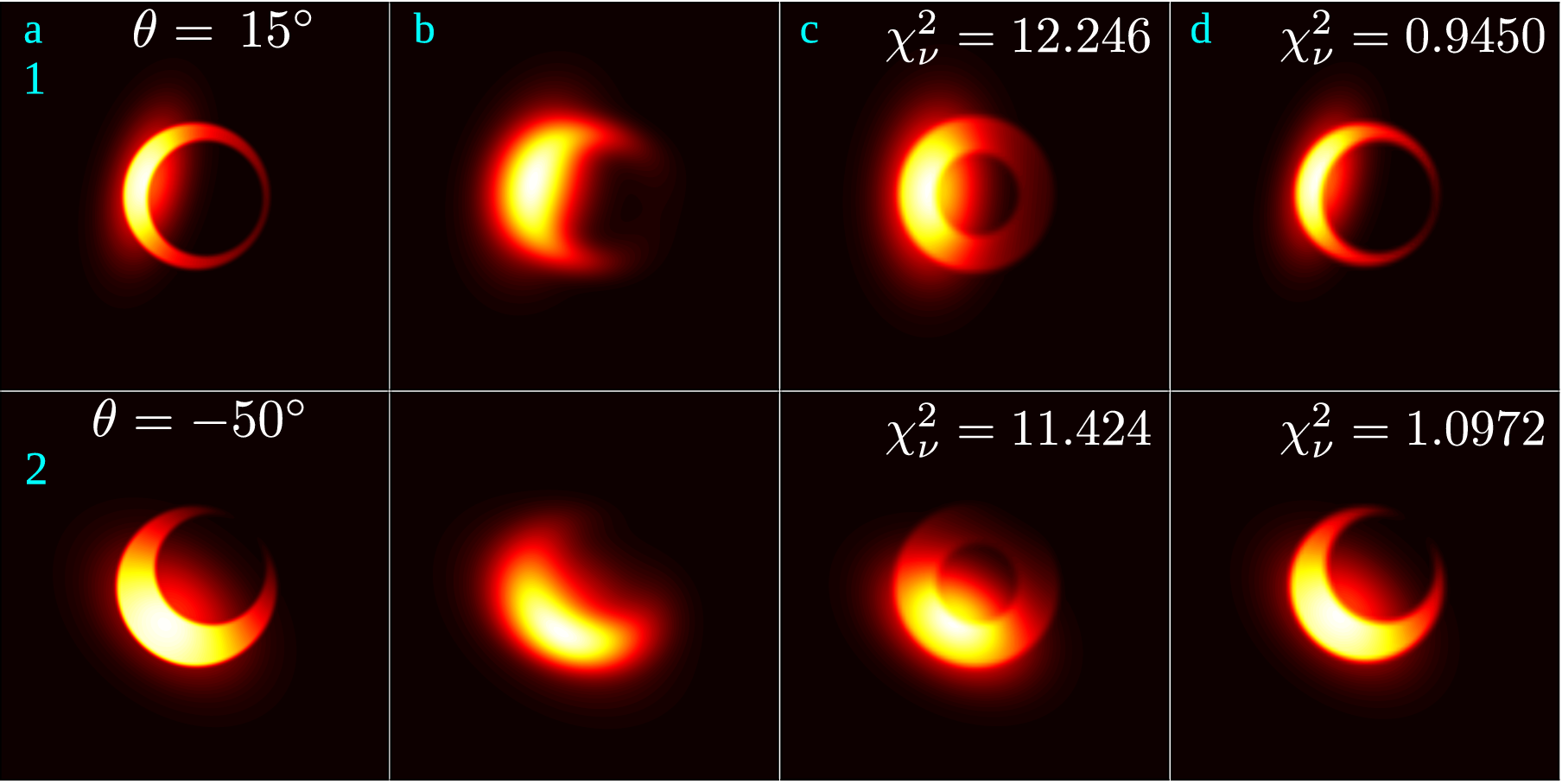}
  \end{center}
\caption{\small Results of the 9-parameter model fitting to the images generated by the 9-parameter model itself. Columns: (a) original image; (b) scattered image; (c) 9-parameter model images fitted to raw data; (d) 9-parameter model images fitted to descattered data. The model parameters are given in Tab.~\ref{tab_model_model}.
\label{fig_model_model}}
\end{figure*}

In order to test the reliability of overall simulation pipeline and especially the fitting software, a few images of the 9-parameter model with arbitrary parameters were generated and scattered, and simulated observational data were produced using MAPS. Subsequent MCMC fitting was able to recover estimates of the model parameters that are very close to the input values, with reduced $\chi^2 \approx 1$ (Fig.~\ref{fig_model_model} and Tab.~\ref{tab_model_model}). We can conclude that the fitting works properly and any $\chi^2_\nu$ significantly greater than unity should be attributed to two major factors: uncertainties in the observation data and inadequacy (excessive simplicity) of the geometric model.

\begin{deluxetable*}{clccccccccccc}[ht!]
  \tablecaption{Fitted 9 parameters to 9-parameter model images in 
                Fig.~\ref{fig_model_model} \label{tab_model_model}}
  \tablewidth{0pt}
  \tablehead{ 
    \colhead{Row} & \colhead{Column} & \colhead{$Z_\mathrm{sp}$} &
    \colhead{$\rex$ } & \colhead{$r_q$ } & \colhead{ecc} & \colhead{fade} & \colhead{$g_\mathrm{ax}$} & 
    \colhead{$a_\mathrm{q}$} & \colhead{$g_\mathrm{q}$} & \colhead{$\theta^\circ$} & 
    \colhead{$\chi^2$} & \colhead{$\chi^2_\nu$} }
  \startdata
  & (a):source & 2.5 & 32 & 0.8 & 0.6 & 0.2 & 1.5 & 0.5 & 0.6 & 15 &  & \\ 
  \vspace{-0.15cm}
   1 & & & & & & & & & & & & \\
  & (d):fitted & 2.50 & 31.8 & 0.81 & 0.82 & 0.41 & 1.49 & 0.51 & 0.60 & 15.43 & 597.2 & \bf{0.945} \\
  \hline   \\[-1.5mm]
  & (a):source & 2.5 & 35 & 0.7 & 0.9 & 0.0 & 1.5 & 0.7 & 0.5 & -50 &  & \\    
  \vspace{-0.15cm}
   2 & & & & & & & & & & & & \\
  & (d):fitted & 2.49 & 34.3 & 0.72 & 1.0 & 0.19 & 1.48 & 0.70 & 0.52 & -49.09 & 693.5 & \bf{1.097}
  \enddata
\end{deluxetable*}

\subsection{Observations of BJPL2013 Model Images}

Here we shall demonstrate the ability of the xringaus model to lock in on model parameters. First we consider in detail the fitting to a random instance of BJPL2013, the process' MCMC histograms, and agreement between the amplitudes and phases of the observed image and the fitted xringaus model. Next simulations shall show the xringaus model lock in on the series of BJPL2013 with one particular variable changed while others are picked randomly. Namely, 
- the series of BJPL2013 with the inclinations in descending order with, with fixed sky orientation and random spins; 
- the series of BJPL2013 with the sizes/masses in ascending order with random inclinations, orientations and spins;
- the series where a randomly selected BJPL2013 image with certain size/mass, inclination, orientation, and spin is observed at the sky orientations from 0$^\circ$ to 360$^\circ$ with the 30$^\circ$  steps.
 
\begin{figure*} [ht!]
\epsscale{0.8}
\plottwo{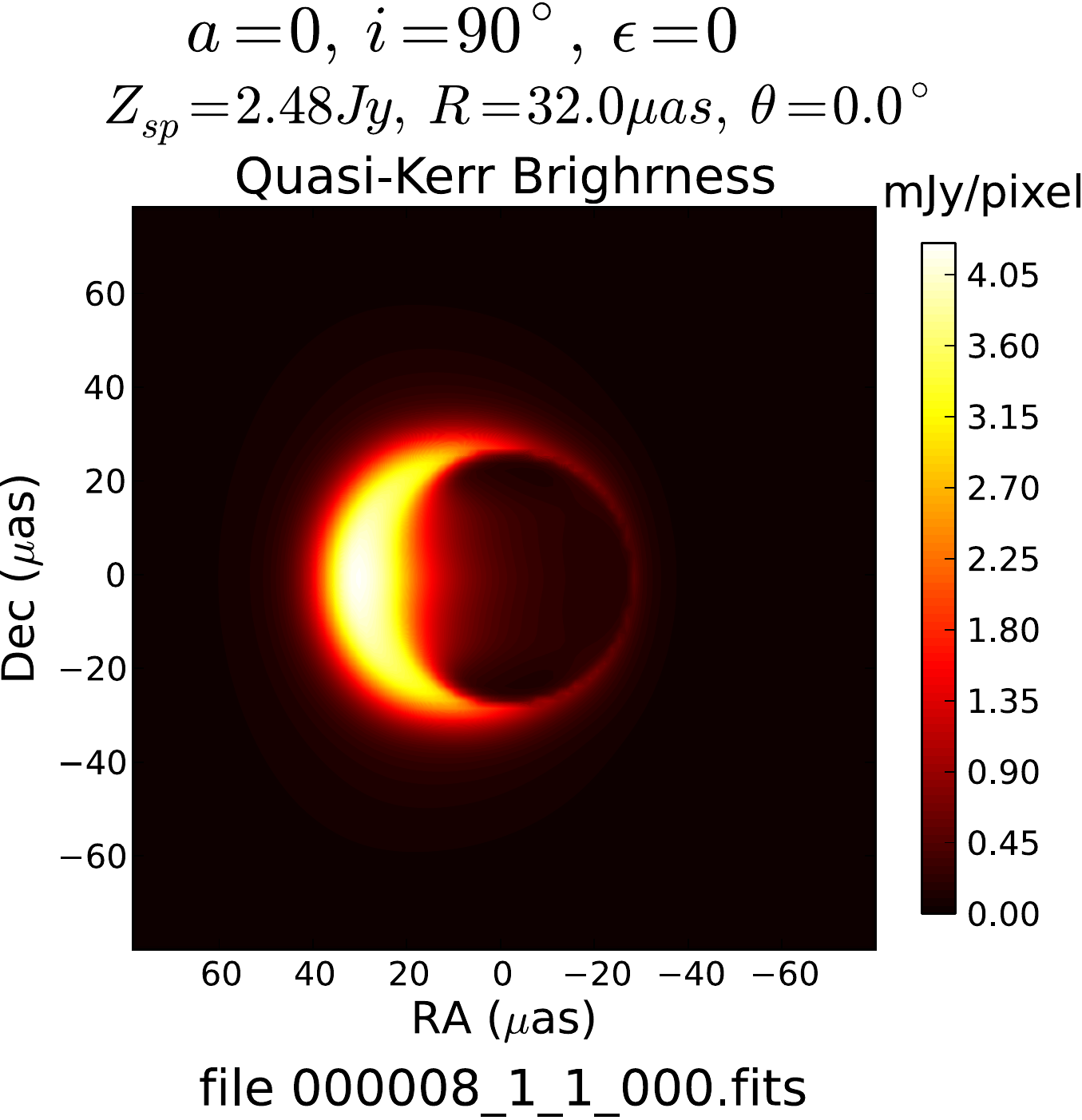}{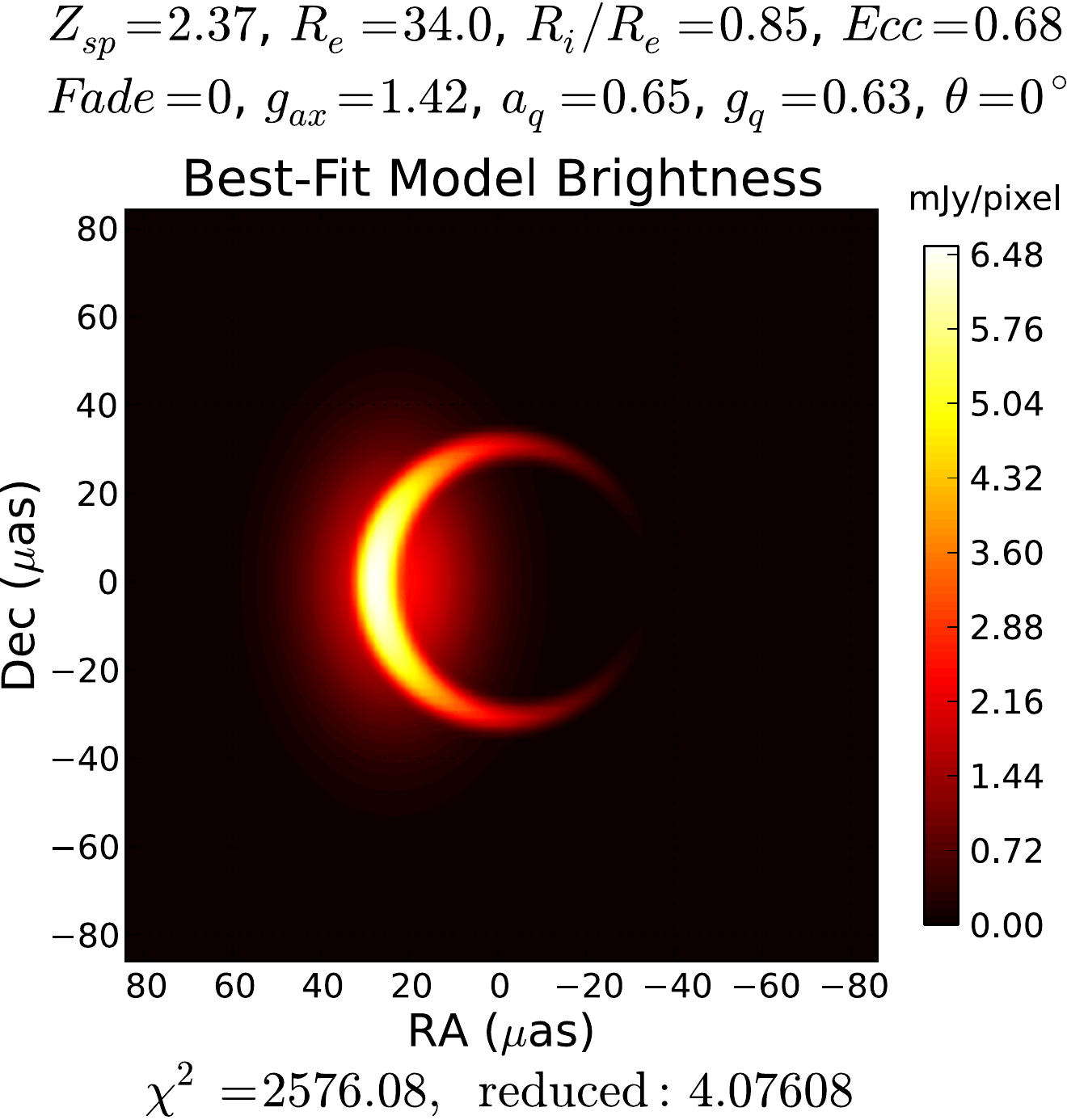}
\caption{\small Comparison of the observed BJPL2013 image on the left and the 9-parameter model fitted to the simulated VLBI observation data on the right. The parameters of both source image and its model fit are printed above the images. 
\label{imag_000008}}
\end{figure*}

As a first example we fit the 9-parameter model to the observation data of a BJPL2013 image with the spin $a=0 \, \mathrm{M}$ and $i=90^\circ$, shown on the left of Fig.~\ref{imag_000008}. The image has been ``scattered" by convolving it with the two-dimensional elliptical Gaussian kernel. 

\begin{figure*}[ht!]
  \begin{center}
        \pdfimageresolution=310
    \includegraphics{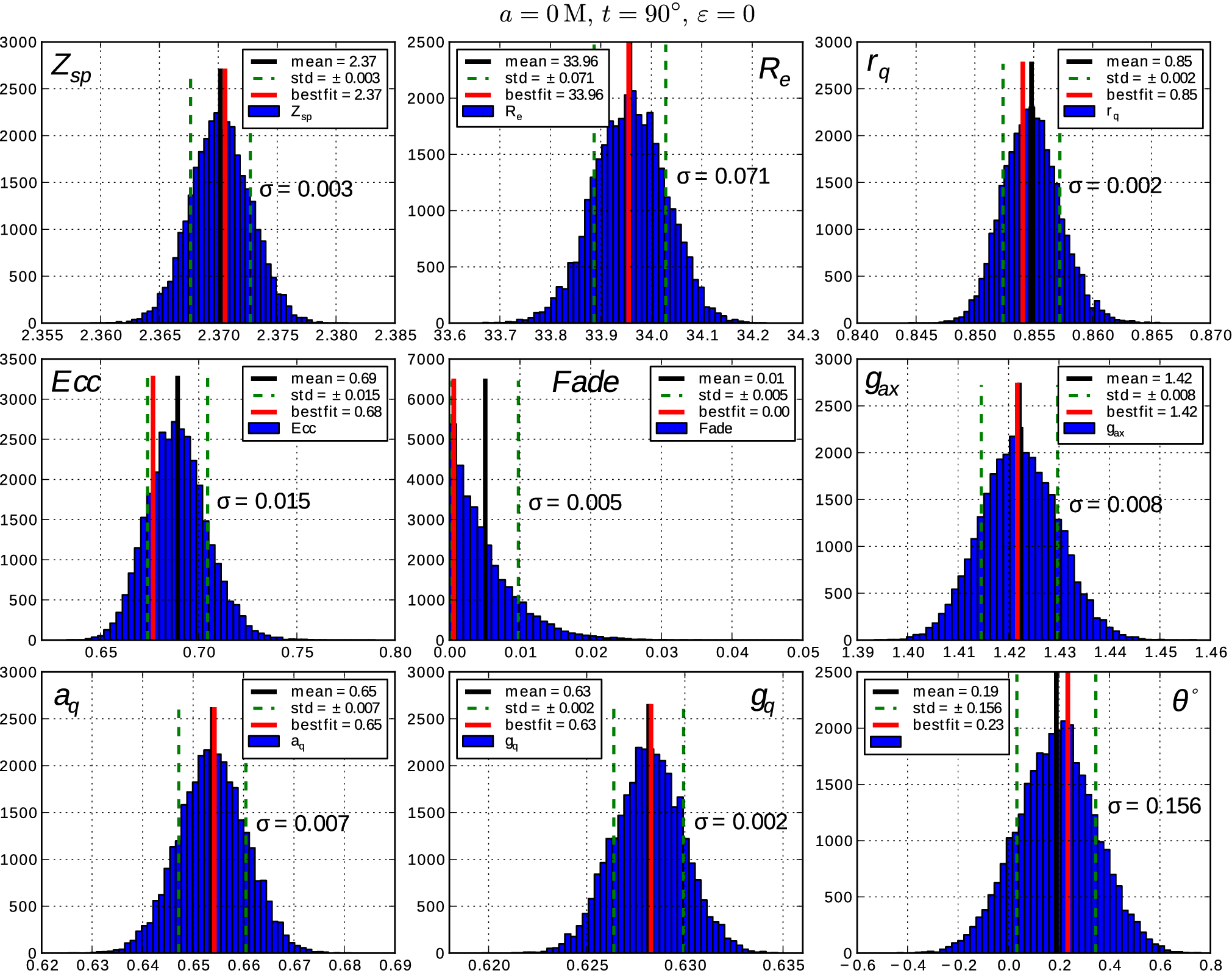}
  \end{center}
\caption{\small Histograms of the Markov Chain Monte-Carlo (MCMC) sequences used in the 9-parameter model fit to simulated data of the Sgr A* BJPL2013 image with the parameters of spin $a=0$ M, inclination $i=90^\circ$, and the residual quadrupole moment parameter $\epsilon=0$. 
\label{hist_000008snd}}
\end{figure*}

\begin{figure*} [ht!]
  \begin{center}
        \pdfimageresolution=330
    \includegraphics{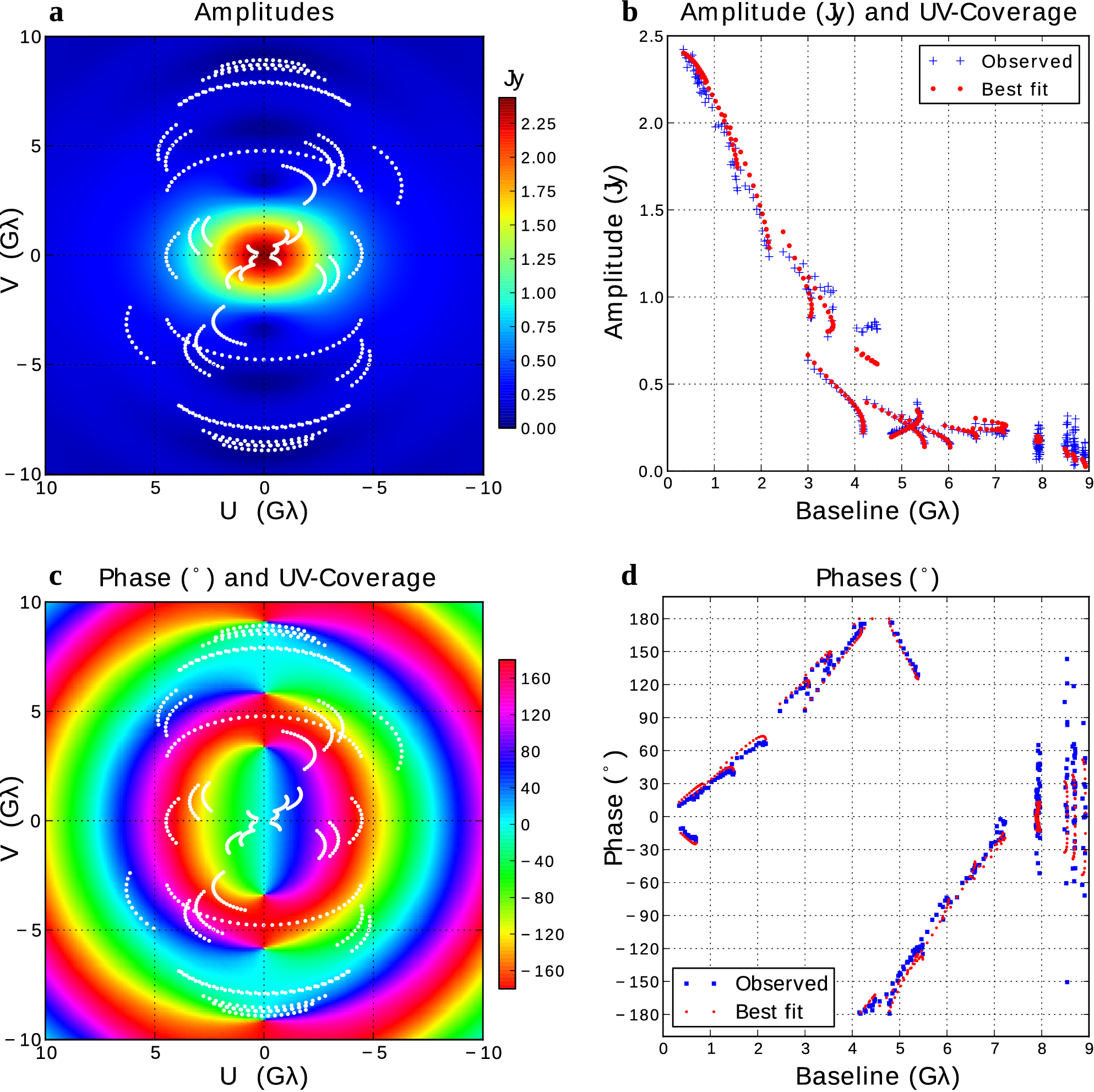}
  \end{center}
\caption{\small Left panels: the simulated VLBI $uv$-coverage is printed as white dots over the 9-parameter model visibility amplitude (a) and phase (c).  Right panels: visibility amplitudes (b) and phases (d) versus baseline lengths. Both observation results (blue pluses for amplitudes and squares for phases) and 9-parameter model values (red dots) are shown.  
\label{vamp_vpha}}
\end{figure*}

The 24-hour observation of the image has been simulated to provide over 300 complex visibility values on the $uv$ plane. Their locations are shown in panels (a,c) of Fig.~\ref{vamp_vpha} as white dots. To remove the effects of scattering the $uv$-data were ``descattered" by multiplication by the inverse of the Fourier transform of the scattering kernel. The MCMC histograms for all nine parameters are shown in Fig.~\ref{hist_000008snd}. All the histograms are narrow and have well-defined maxima. Values of the standard deviations $\sigma$ characterize the errors at $\sim 1\%-2\%$.

In order to compare the BJPL2013 image and synthesized 9-parameter model images, they are juxtaposed on Fig.~\ref{imag_000008}. The dark shadow areas are of comparable sizes; therefore, the model can be used to estimate the size of the black hole shadow. Note that the source image and the fitted 9-parameter model have close outer radii (32 and 34 $\mu$as), sizes of the shadows, and the same orientation ($\theta = 0^\circ$ for both).  The MCMC fitting procedure applied to several other source images with different $(a,t)$ parameter combinations produces qualitatively similar histograms.

The 9-parameter best fit model visibility amplitude and phase are imaged in panels (a,c) in Fig.~\ref{vamp_vpha}. The white dots of the $uv$-coverage show the sampling points where differences between the observations and the model were minimized. Panels (b,d) compare the 9-parameter model visibility amplitudes and phases with those observed.

\begin{figure*} [ht!]
  \begin{center}
        \pdfimageresolution=200
    \includegraphics{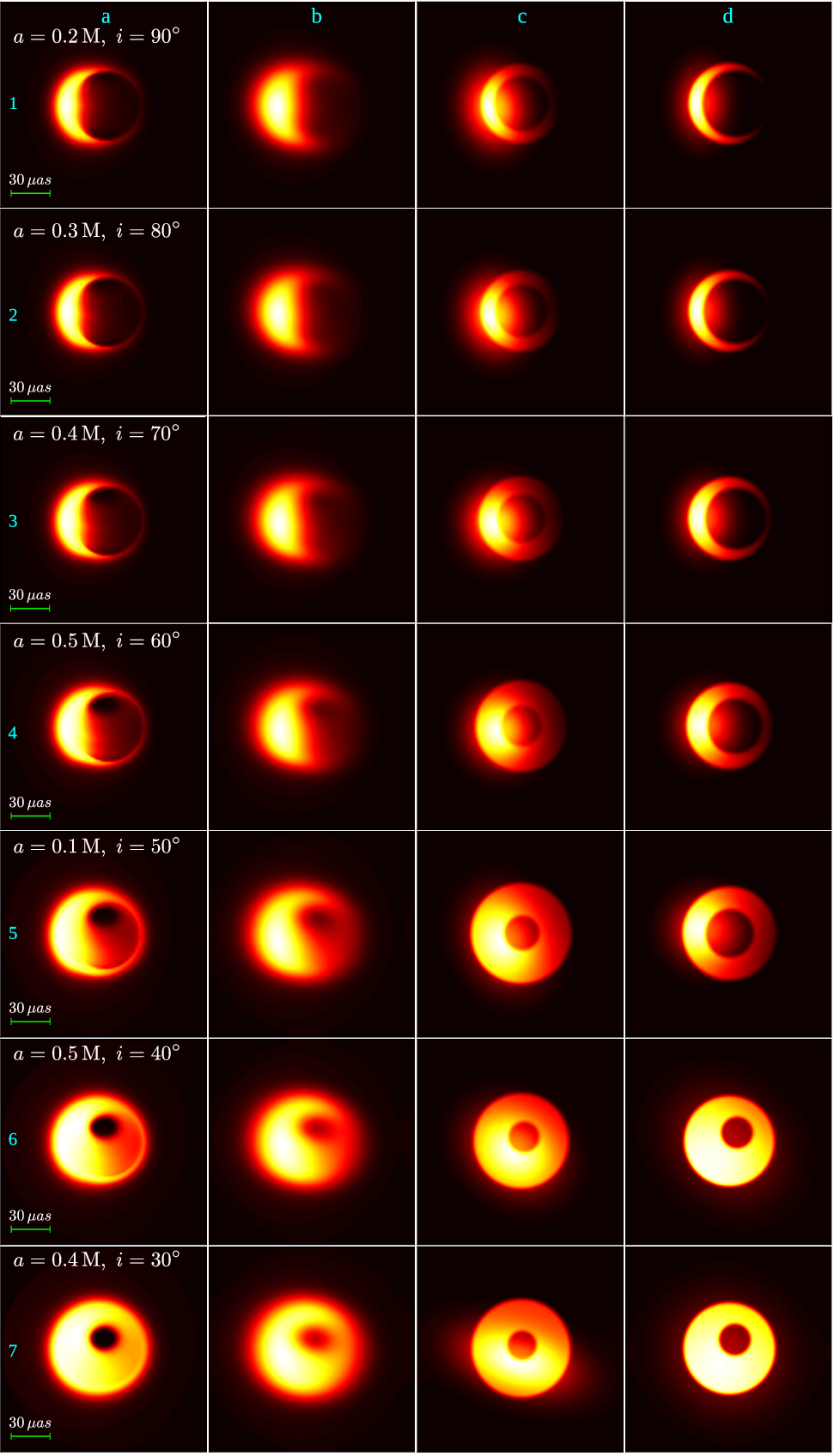}
  \end{center}
\caption{\small Comparison of the observed Sgr A* images and the 9-parameter model images fitted to the VLBI simulated observation data. The images have been selected with random $a$ parameter and with the inclination angle changing from $i=90^\circ$ at the top to $i=30^\circ$ at the bottom. They are presented in column (a). Column (b) comprises the same image exposed to scattering by convolving the original images with the 2D Gaussian kernel. Columns (c,d) feature the fitted 9-parameter model images. Column (c) is made of the model fits to the non-descattered, and (d) - to descattered observation data. The fitted model parameters are given in Tab.~\ref{tab_inclin}. 
\label{fig_compar_incl}}
\end{figure*}

The next series of simulations provides several examples of fitting the 9-parameter model to the BJPL2013 images of black hole with different spin inclination angles $i$, changing from 90$^\circ$ to 30$^\circ$. The results for both non-descattered and descattered observation data are shown in Fig.~\ref{fig_compar_incl}. The best-fit model parameters are listed in Tab.~\ref{tab_inclin}. The descattering apparently improves the fitted image quality. The worst quality is at $i=50^\circ$, because the 9-parameter model is designed to imitate either the ``crescent" view of the edge-on spin ($90^\circ-70^\circ$), or the ``funnel" view of the face-on spin ($40^\circ-30^\circ$), while $i=50^\circ$ is in between the two states. The fitted model parameters will be used to estimate those of the black hole.

\begin{deluxetable*}{ccccccccccccc} [ht!]
  \tablecaption{ Fitted parameters for various inclinations in Fig.~\ref{fig_compar_incl} \label{tab_inclin}}
  \tablewidth{0pt}
  \tablehead{ 
    \colhead{$i^\circ$} & \colhead{$a$ (M)} & \colhead{$Z_\mathrm{sp}$} &
    \colhead{$\rex$ } & \colhead{$\rin$ } & \colhead{ecc} & \colhead{fade} & \colhead{$g_\mathrm{ax}$} & 
    \colhead{$a_\mathrm{q}$} & \colhead{$g_\mathrm{q}$} & \colhead{$\theta^\circ$} & 
    \colhead{$\chi^2$} & \colhead{$\chi^2_\nu$} }
  \startdata
  90 & 0.2 & 2.38 & 33.9 & 28.5 & 0.91 & 0.02 & 1.29 & 0.70 & 0.61 &  1.2  & 2311.6 & 3.6575 \\
  80 & 0.3 & 2.39 & 33.8 & 27.7 & 0.80 & 0.08 & 1.25 & 0.73 & 0.61 &  0.9  & 2095.3 & 3.3153 \\
  70 & 0.4 & 2.38 & 34.7 & 26.2 & 0.56 & 0.08 & 1.19 & 0.79 & 0.58 &  1.2  & 2605.5 & 4.1226 \\
  60 & 0.5 & 2.39 & 36.0 & 23.5 & 0.36 & 0.11 & 1.12 & 0.92 & 0.54 &  2.3  & 3988.2 & 6.3104 \\
  50 & 0.1 & 2.39 & 39.2 & 20.6 & 0.00 & 0.18 & 1.05 & 1.05 & 0.46 &  8.8  & 5434.6 & 8.5991 \\
  40 & 0.5 & 2.47 & 37.9 & 13.9 & 0.38 & 0.51 & 1.99 & 0.80 & 0.37 & -48.  & 7303.2 & 11.556 \\
  30 & 0.4 & 2.47 & 38.7 & 13.3 & 0.33 & 0.73 & 2.00 & 0.93 & 0.34 & -55.  & 3279.0 & 5.1883
  \enddata
\end{deluxetable*}

\begin{figure*}  [ht!]
  \begin{center}
        \pdfimageresolution=350
    \includegraphics{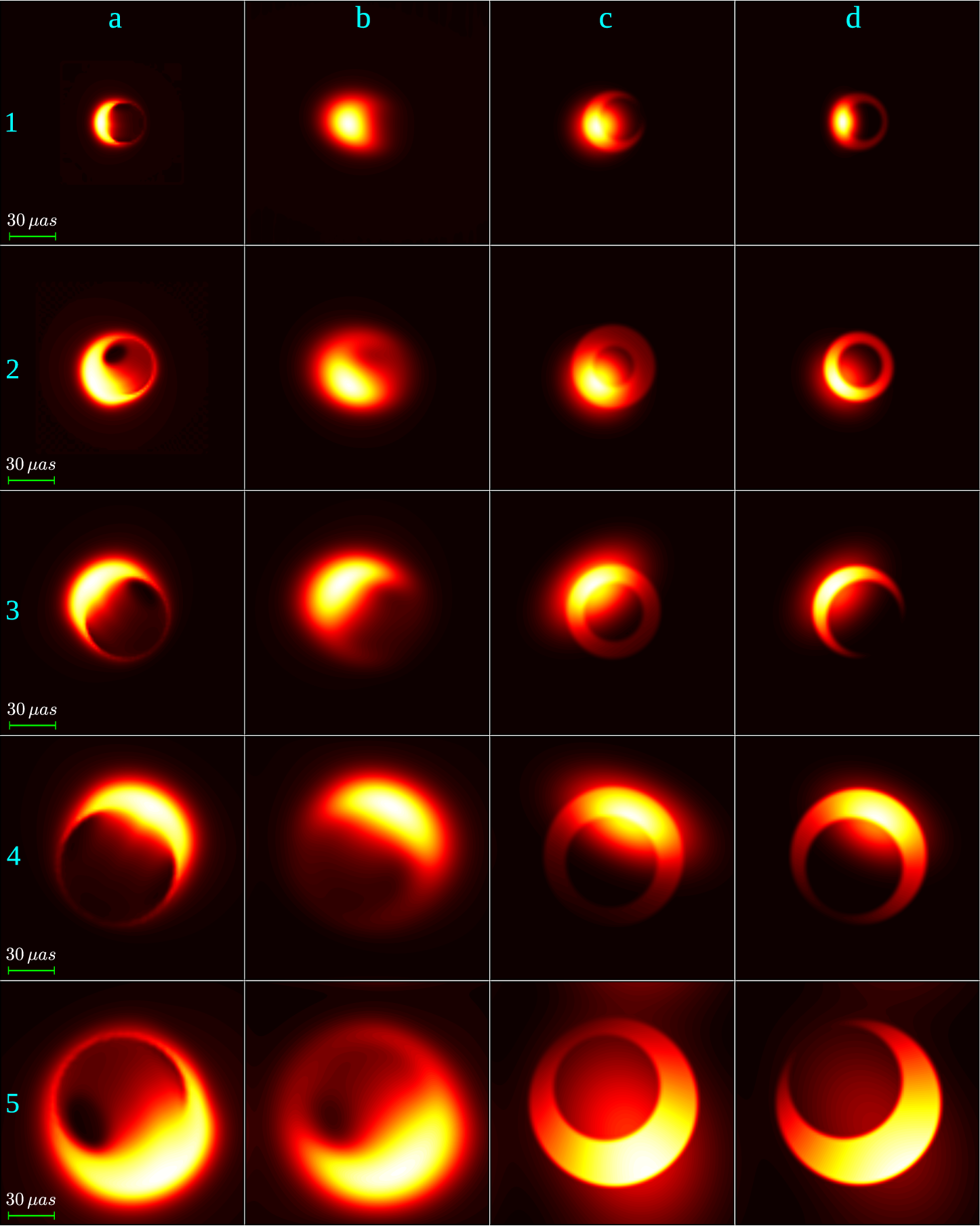}
  \end{center}
\caption{\small A number of fitting examples for various black hole diameters (i.e. masses) and orientations. The spins and inclinations are chosen randomly. Columns:(a) original image; (b) scattered image; (c) 9-parameter model images fitted to raw data; (d) 9-parameter model images fitted to descattered data.  The fitted model parameters are listed in Tab.~\ref{tab_diam}.
\label{fig_model9prm_n_5zooms_3rows}}
\end{figure*}

Estimation of the black hole shadow size is instrumental in measurement its mass. Fig.~\ref{fig_model9prm_n_5zooms_3rows} illustrates the 9-parameter model fitting results over a wide range of the black hole radii from 16 $\mu$as to 51.2 $\mu$as. The numerical values of best-fit models are placed in Tab.~\ref{tab_diam}. Not only the diameters, but also the orientations of the source images vary, and the fitted model images follow both the sizes and the orientations. Again, the quality is higher if the $uv$-data have been descattered before fitting the 9-parameter model.

\begin{deluxetable*}{ccccccccccccccccc}  [ht!]
  \tablecaption{Fitted parameters for various diameters and orientations in
                Fig.~\ref{fig_model9prm_n_5zooms_3rows} \label{tab_diam}}
  \tablewidth{0pt}
  \tablehead{ 
    \colhead{Row} &
    \colhead{$\theta^\circ_\mathrm{orig}$} & \colhead{$R_{\rm orig}/32$} & \colhead{$R_{\rm orig}$} &
    \colhead{$i^\circ$ (M)} & \colhead{$a$} & \colhead{$Z_\mathrm{sp}$} &
    \colhead{$\rex$ } & \colhead{$\rin$ } & \colhead{ecc} & \colhead{fade} & \colhead{$g_\mathrm{ax}$} & 
    \colhead{$a_\mathrm{q}$} & \colhead{$g_\mathrm{q}$} & \colhead{$\theta^\circ$} & 
    \colhead{$\chi^2$} & \colhead{$\chi^2_\nu$} }
  \startdata
  1 &    0 & 0.5 & 22.4 & 80 & 0.3 & 2.47 & 20.2 & 13.8 & 0.56 & 1.00 & 1.00 & 0.67 & 0.72 & -1.5 & 2009.9 & 3.1802 \\
  2 &  -30 & 0.7 & 12.8 & 50 & 0.4 & 2.45 & 24.3 & 16.3 & 0.16 & 0.44 & 1.37 & 0.87 & 0.60 & -39  & 5876.8 & 9.2988 \\
  3 &   45 & 1.0 & 32.0 & 70 & 0.4 & 2.38 & 34.7 & 26.2 & 0.56 & 0.08 & 1.19 & 0.80 & 0.58 & 1.2  & 2605.5 & 4.1226 \\
  4 &  120 & 1.4 & 44.8 & 90 & 0.4 & 2.32 & 48.4 & 34.9 & 0.55 & 0.04 & 1.10 & 0.66 & 0.55 & 116  & 3244.5 & 5.1337 \\
  5 & -120 & 1.6 & 51.2 & 60 & 0.0 & 2.38 & 58.7 & 40.7 & 0.96 & 0.00 & 1.41 & 1.47 & 0.49 & -124 & 4913.5 & 7.7746
  \enddata
\end{deluxetable*}

\begin{figure}   [ht!]
\plotone{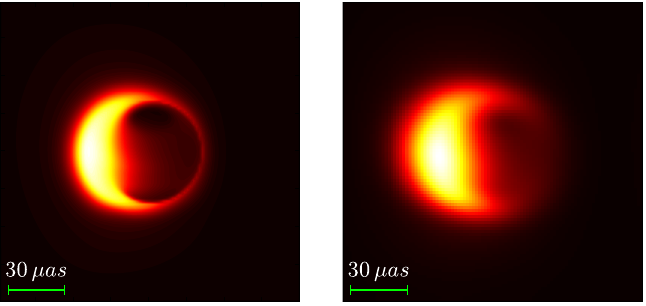}
\caption{\small An ideal BJPL2013 image selected for observing in different orientations (left) and the same one scattered (right). The spin is zero, and the spin inclination is $70^\circ$.
\label{fig_000208_2col}}
\end{figure}

\begin{figure*}[ht!]
  \begin{center}
    \pdfimageresolution=350
    \includegraphics{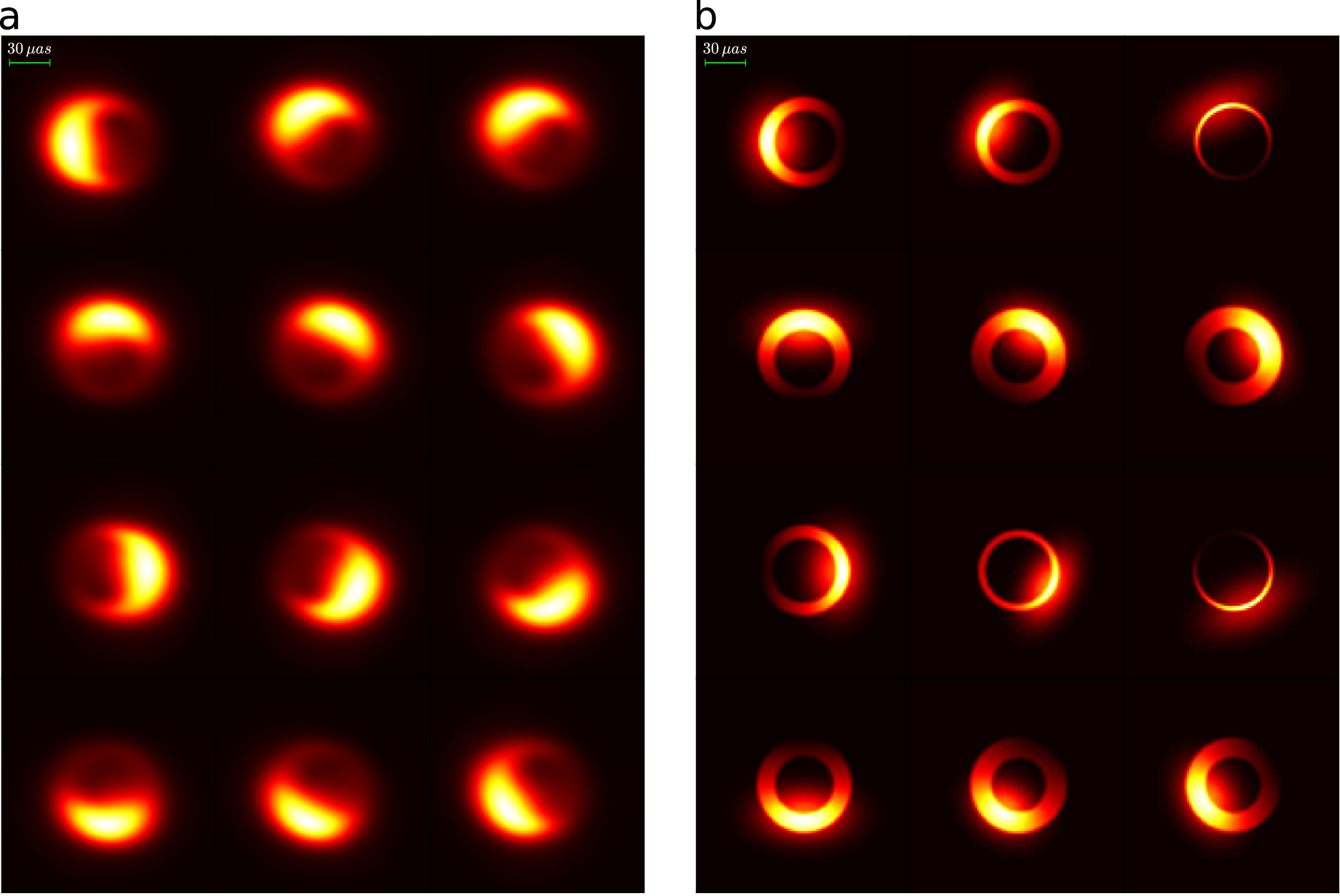}
  \end{center}
\caption{\small 9-parameter model fitting results for all possible orientations of the observed black hole $30^\circ$ apart. (a) scattered ideal images; (b) model fitted to the descattered UV-data. The fitted model parameters are provided in Tab.~\ref{tab_prmori}.
\label{fig_compar_ori_2col_3x4}}
\end{figure*}

In order to show how well the 9-parameter model fitted image follows the angle of the source image orientation we made a series of 12 simulated observations of the same BJPL2013 image in different orientations, from $\theta = 0^\circ$ to $\theta = 330^\circ$ with the step $\Delta\theta = 30^\circ$. The original BJPL2013 image (spin $a = 0$ M, inclination $i = 70^\circ$) and its scattered view are shown in Fig.~\ref{fig_000208_2col}. The resulting sequence of the 9-parameter model fits made to the descattered observation data is shown in panel (b) of Fig.~\ref{fig_compar_ori_2col_3x4}. The major numerical values of the fitted parameters are given in Tab.~\ref{tab_prmori}.

\begin{deluxetable}{cccccccc}  [ht!]
  \tablecaption{ Fitted parameters for various orientations in Fig.~\ref{fig_compar_ori_2col_3x4}
                 \label{tab_prmori}}
  \tablewidth{0pt}
  \tablehead{ 
    \colhead{$\theta^\circ_{\mathrm{orig}}$} & \colhead{$\theta^\circ$ } & \colhead{$R_{\mathrm{orig}}$ } &
    \colhead{$\rex$ } & \colhead{$\rin$ } & \colhead{$Z_\mathrm{sp}^\mathrm{orig}$ } & 
    \colhead{$Z_\mathrm{sp}$} & \colhead{$\chi^2_\nu$ } }
  \startdata
  0   & 3   & 32.0 & 36.2 & 25.3 & 2.48 & 2.35 & 4.01  \\ 
  30  & 29  & 32.0 & 34.3 & 25.0 & 2.48 & 2.37 & 6.43  \\
  60  & 66  & 32.0 & 29.7 & 29.7 & 2.48 & 2.41 & 12.8  \\
  90  & 89  & 32.0 & 37.5 & 24.4 & 2.48 & 2.41 & 6.03  \\
  120 & 120 & 32.0 & 38.3 & 22.6 & 2.48 & 2.37 & 5.41  \\
  150 & 151 & 32.0 & 39.8 & 22.3 & 2.48 & 2.37 & 5.23  \\
  180 & 182 & 32.0 & 36.4 & 26.2 & 2.48 & 2.36 & 8.29  \\
  210 & 207 & 32.0 & 32.5 & 27.3 & 2.48 & 2.38 & 9.95  \\
  240 & 245 & 32.0 & 31.1 & 30.8 & 2.48 & 2.41 & 16.4  \\
  270 & 270 & 32.0 & 38.2 & 23.7 & 2.48 & 2.43 & 7.72  \\
  300 & 300 & 32.0 & 39.0 & 22.2 & 2.48 & 2.37 & 4.54  \\
  330 & 331 & 32.0 & 38.2 & 22.5 & 2.48 & 2.36 & 5.58 
  \enddata
\end{deluxetable}

\subsection{Observations of GRMHD Model Images}

\begin{figure*}  
\begin{center}
    \pdfimageresolution=350
    \includegraphics{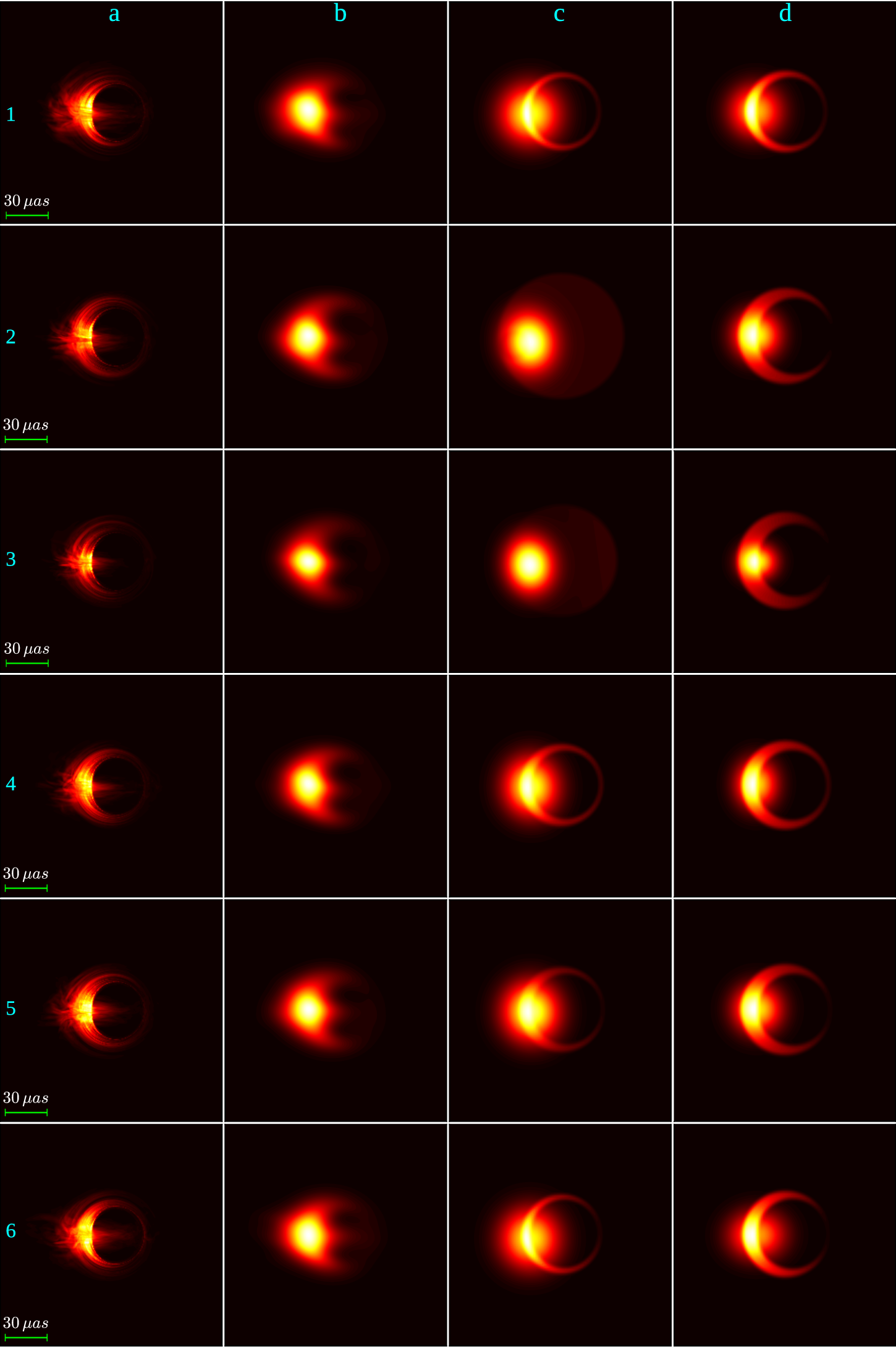}   
\end{center}
\caption{\small Results of the 9-parameter model fitting to frames from the Sgr A* dynamics simulation movie. Columns: (a) original image; (b) scattered image; (c) 9-parameter model images fitted to raw data; (d) 9-parameter model images fitted to descattered data. The model parameters are given in Tab.~\ref{tab_dolence9prm}.
\label{fig_dolence6_9prm}}
\end{figure*}

So far we have used for observation the BJPL2013 Sgr A* model images provided by A. Broderick. However, there are other black hole accretion flow models. Here we consider fitting the nine-parameter model to the model images created by M.~Moscibrodzka and J.~Dolence \citep{Monika2012galacenter}. Several observed images making the leftmost column (a) in Fig.~\ref{fig_dolence6_9prm} are randomly selected frames from a whole 23-hour ``movie" simulating Sgr A* accretion flow in dynamics. One can notice that the fitting is not always successful: the inner radius of two fits is about zero. However, descattering solves this problem and improves the model image quality, as seen in column (d). The numerical results of the model fits are shown in Tab.~\ref{tab_dolence9prm}.

\begin{deluxetable*}{ccccccccccccc}  [ht!]
  \tablecaption{Fitted 9 parameters for several movie frames in
                Fig.~\ref{fig_dolence6_9prm} \label{tab_dolence9prm}}
  \tablewidth{0pt}
  \tablehead{ 
    \colhead{Row} & \colhead{Frame} & \colhead{$Z_\mathrm{sp}$} &
    \colhead{$\rex$ } & \colhead{$\rin$ } & \colhead{ecc} & \colhead{fade} & \colhead{$g_\mathrm{ax}$} & 
    \colhead{$a_\mathrm{q}$} & \colhead{$g_\mathrm{q}$} & \colhead{$\theta^\circ$} & 
    \colhead{$\chi^2$} & \colhead{$\chi^2_\nu$} }
  \startdata
   1 & 03634 & 2.80 & 31.6 & 27.2 & 0.75 & 0.35 & 0.98 & 1.02 & 0.76 &  0.5 & 10684.5 & 16.9059 \\
   2 & 04356 & 2.54 & 37.2 & 30.1 & 1.00 & 0.43 & 0.76 & 1.01 & 0.68 &  1.4 &  9611.4 & 15.2079 \\
   3 & 05818 & 1.81 & 37.8 & 28.6 & 0.70 & 0.00 & 0.55 & 1.09 & 0.55 & -2.4 &  4371.1 &  6.9163 \\
   4 & 08768 & 2.71 & 34.4 & 27.7 & 0.68 & 0.32 & 0.85 & 0.86 & 0.65 &  1.4 &  8700.9 & 13.7672 \\
   5 & 08970 & 2.66 & 35.3 & 0.80 & 0.83 & 0.18 & 0.82 & 1.01 & 0.68 &  0.9 &  7607.6 & 12.0373 \\
   6 & 09988 & 2.85 & 33.5 & 0.83 & 0.86 & 0.17 & 0.86 & 1.17 & 0.66 & -1.9 &  6426.9 & 10.1692
  \enddata
\end{deluxetable*}

\begin{figure}[ht!]
  \begin{center}
        \pdfimageresolution=350
    \includegraphics{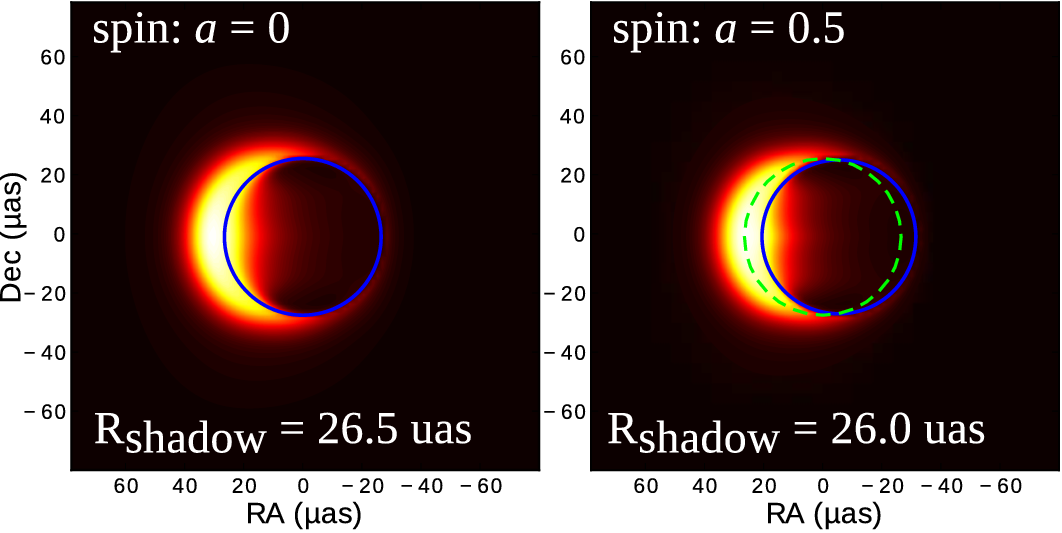}
  \end{center}
\caption{\small Black hole shadow sizes (blue circles) for different spins, $a=0$ on the left and $a=0.5$ on the right. The green dashed circle on the right panel  is drawn for comparison. It has the same size as that of the blue circle on the left panel. The shadow radius slightly decreases with the growing spin.
\label{fig_shadow_size}}
\end{figure}

In the standard General Relativity (GR) framework the black hole shadow must have strictly circular form. We assume no deviations from GR, so we can only try to estimate two of the ``hairless" black hole parameters: its mass and its spin. Knowing the distance to the black hole, the mass is calculated from its diameter. The spin, ranging from $a=0$ to $a=0.998$, affects the shadow size and relative position. For Sgr A* the theoretical shadow radius is calculated as  
\begin{equation}
  \label{r_shadow_theor}
  R_{\mathrm{shadow}} = (4.5 + 0.7\sqrt{1 - a^2})\cdot 5.1 \,\, \mu \mathrm{as} ,
\end{equation}
where $a$ is the spin. This is illustrated in Fig.~\ref{fig_shadow_size}, where the juxtaposition of the BJPL2013 images with the spins $a=0$ and $a=0.5$ is shown. Unfortunately, the shadow diameter decreases very slightly with growing spin, from 26.5 $\mu$as to 26.0 $\mu$as. This small change will most likely be swamped within the error bars in actual measurement. The shift of the shadow circle off the center due to the spin is much more salient, as one can notice in the right panel of Fig.~\ref{fig_shadow_size}, where the shadow position at $a=0$ is outlined with the dashed circle. 

\begin{figure}[ht!]
\plotone{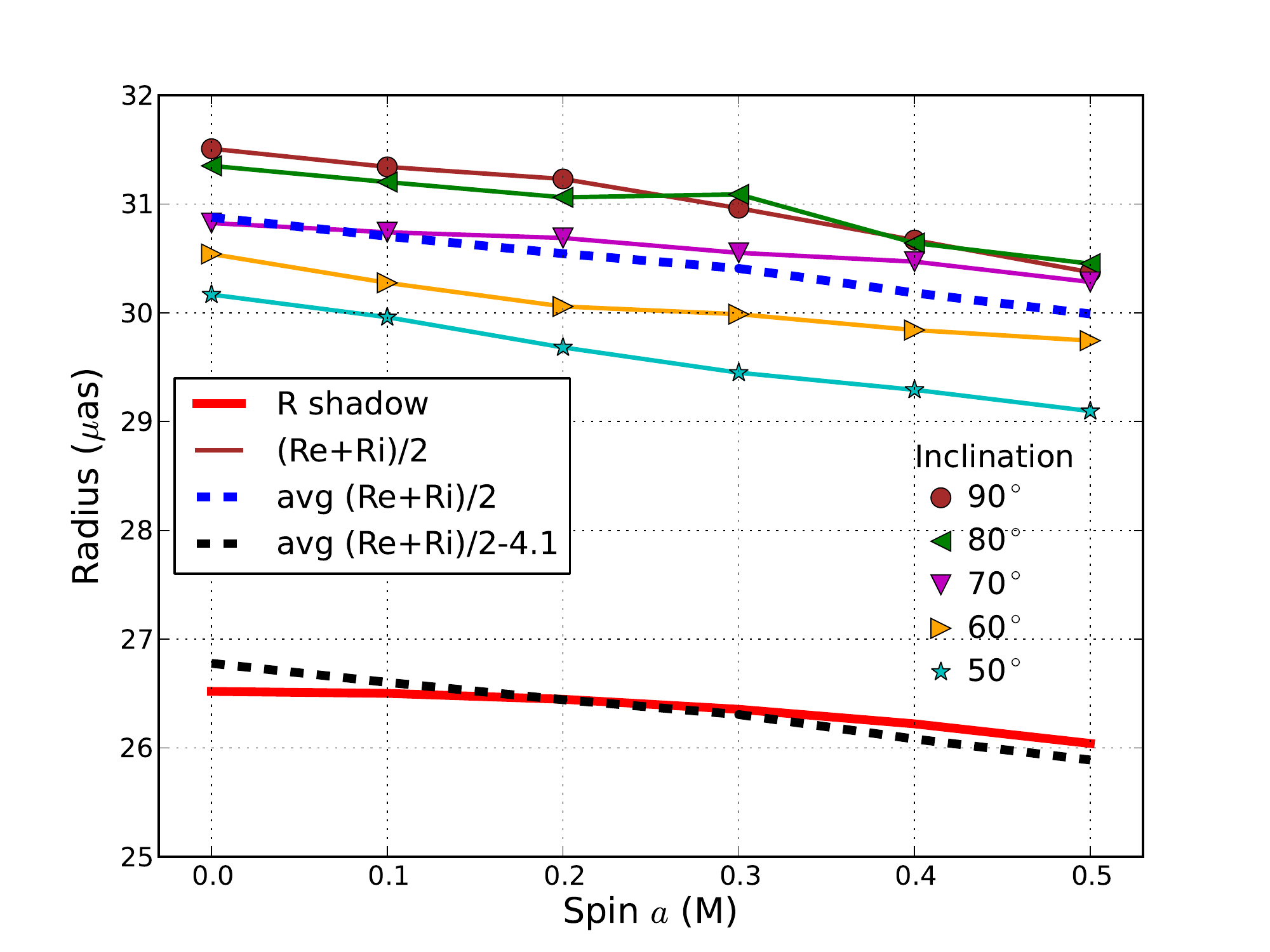}
\caption{\small Shadow radius versus spin for $a=0$ to $a=0.5$ and inclinations from 90$^\circ$ to 50$^\circ$.
\label{fig_Rshadow_Re_Ri_mean_vs_spin}}
\end{figure}

Attempts to assess the true shadow radius from the estimated model parameters led us to the formula 
\begin{equation}
  \label{r_shadow}
  R_{\mathrm{shadow}}=\onehalf(\rex+\rin) - \xi(t),
\end{equation}
where $\xi(t)$ are some empirical values. Fig.~\ref{fig_Rshadow_Re_Ri_mean_vs_spin} explains derivation of \eqref{r_shadow}. Thin lines with markers plot the mean radius, $R_{\mathrm{mean}} = \onehalf(\rex+\rin)$, for different inclinations from $i=90^\circ$ to $i=50^\circ$ as functions of the spin, over the interval from $a=0$ M to $a=0.5$ M. Each of the curves needs to be ``pulled down" by a subtrahend $\xi(t)$ to approximately overlap the thick red curve of the theoretical shadow radius. On average, the subtrahend is shown to be $\xi \approx 4.1$. Its values picked for inclinations from 90$^\circ$ to 50$^\circ$ are presented in Tab.~\ref{rsh_xi}. This method does not seem to provide reliable information on the shadow size at the lower inclinations, 40$^\circ$ to 30$^\circ$, so they are not presented in Tab.~\ref{rsh_xi}. 


\begin{deluxetable}{cccccc}  [ht!]
  \tablecaption{ Subtrahend $\xi(t)$ from $\onehalf(\rex+\rin)$ \label{rsh_xi}}
  \tablewidth{0pt}
  \tablehead{\colhead{$i^\circ$} & \colhead{90} & \colhead{80} & \colhead{70} & \colhead{60} & \colhead{50}} 
  \startdata
    $\xi(t)$ & 4.65  & 4.60 & 4.25 & 3.65 & 3.23 
  \enddata
\end{deluxetable}

\begin{figure}[ht!]
\plotone{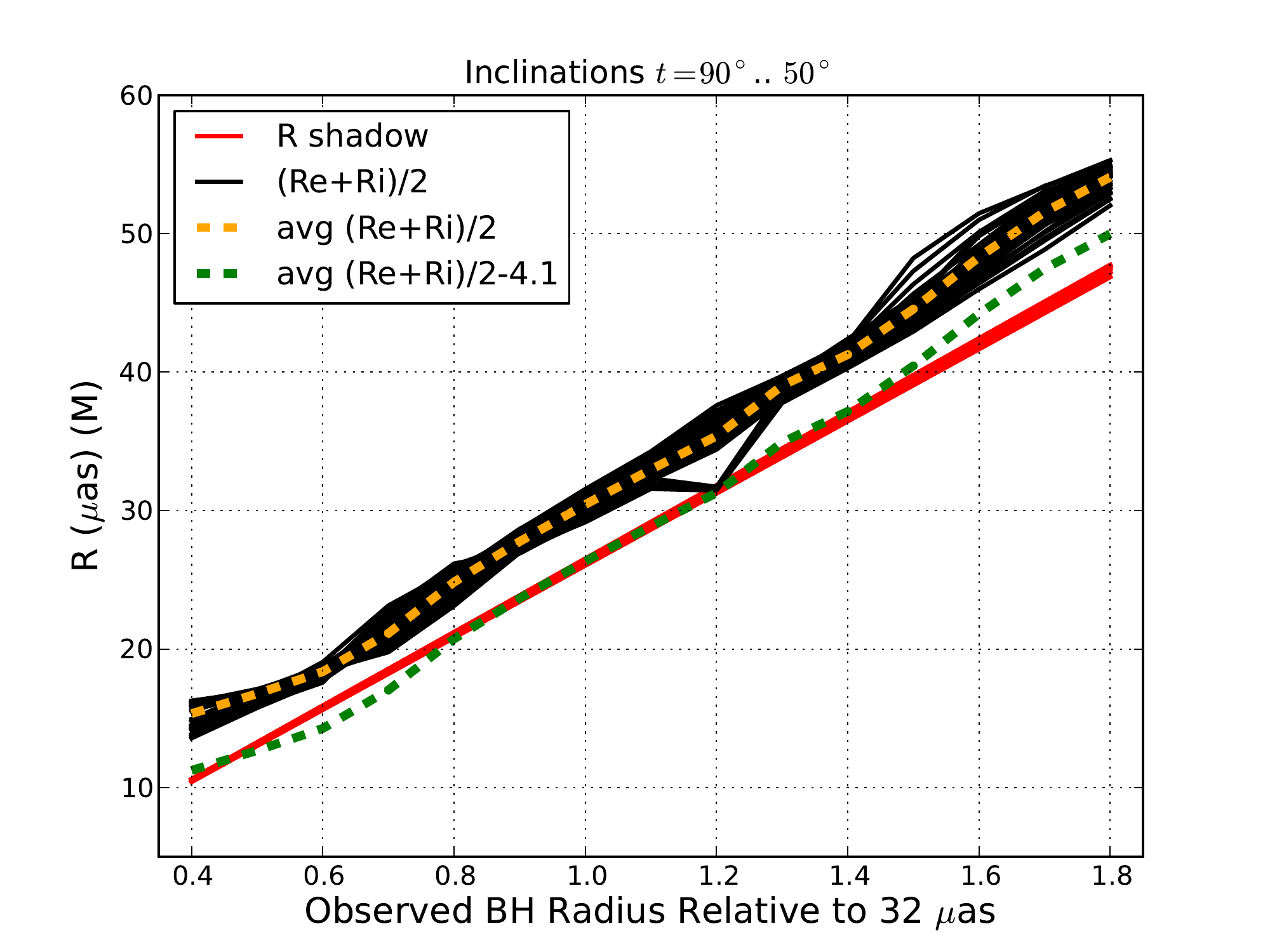}
\caption{\small True shadow radius (red) compared with the mean of the fitted model parameters $R_{external}$ and $R_{internal}$ over the range of the black hole masses observed as the range of their sizes from 0.4 to 1.8 of 32 $\mu$as.
\label{fig_Rshadow_Re_Ri_mean_vs_Rrel_all_inclin}}
\end{figure}

Varying the sizes (and hence, masses) of observed black holes with different spins and at different inclinations shows that \eqref{r_shadow} works well. In Fig.~\ref{fig_Rshadow_Re_Ri_mean_vs_Rrel_all_inclin} the mean model radii $\onehalf(\rex+\rin)$ plotted as thin black curves for all possible spins and inclinations merge into a thick bundle. The average over them all is shown as a dashed yellow line. Again, as in the previous dependence in Fig.~\ref{fig_Rshadow_Re_Ri_mean_vs_spin}, the true observed shadow size can be approximated by the formula  $R_{\mathrm{shadow}}=\onehalf(\rex+\rin) - 4.1$. 

\begin{figure}[ht!]
\plotone{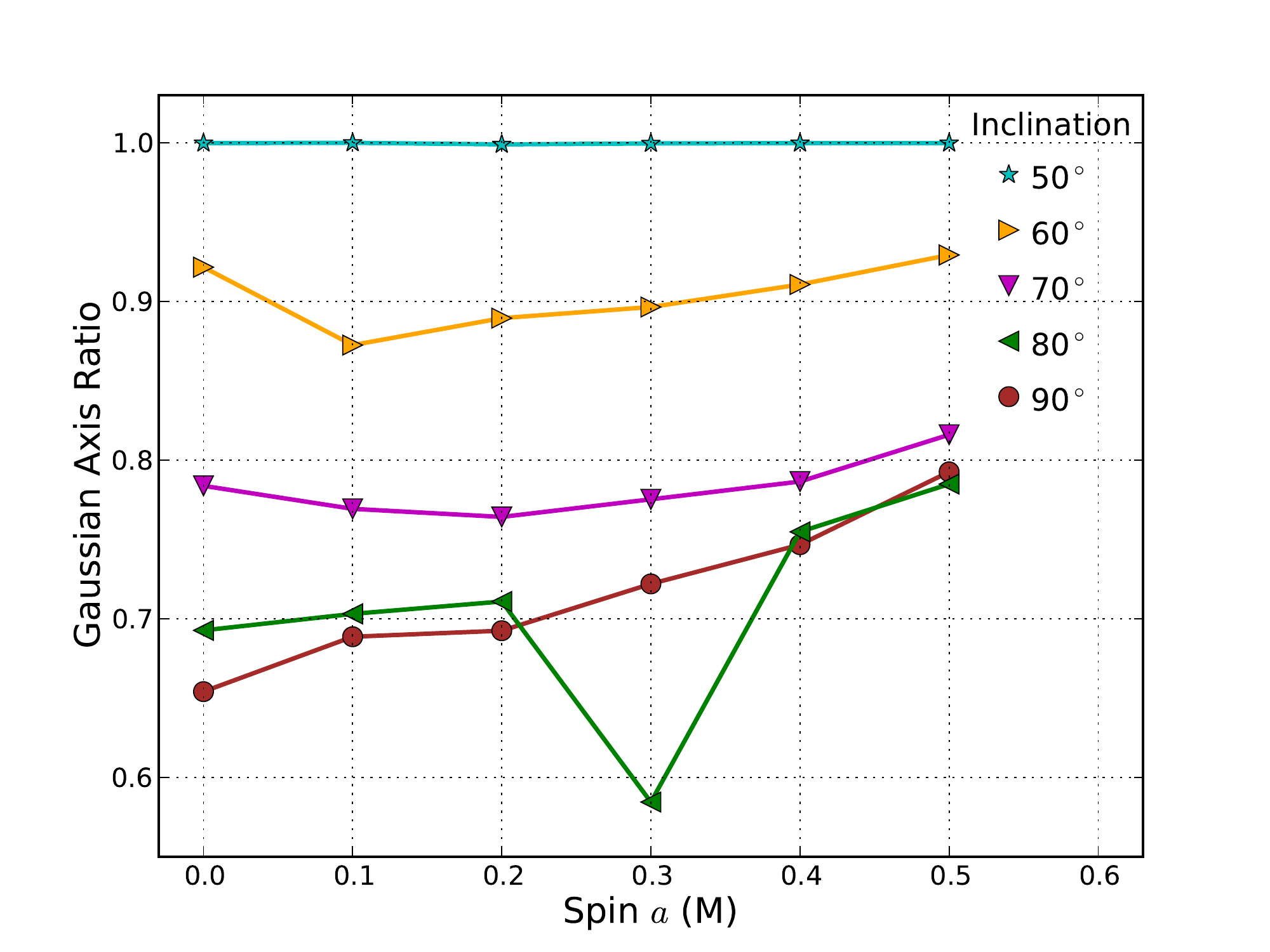}
\caption{\small The Gaussian axes ratio versus spin for $a=0$ to $a=0.5 $and inclinations from 90$^\circ$ to 50$^\circ$.
\label{fig_Gaussian_aq_vs_spin}}
\end{figure}

The elliptical Gaussian in the 9-parameter model highlights the thicker part of the crescent. The ellipse is near circular if its axes ratio $a_q \approx 1$, and becomes thinner with $a_q$ decreasing. Fig.~\ref{fig_Gaussian_aq_vs_spin} shows dependence of $a_q$ on the spin $a$. For the accretion flow positions close to edge-on the Gaussian grows thicker with the spin. For lower inclinations the dependence is more complex. 

\begin{figure*}[ht!]
\plottwo{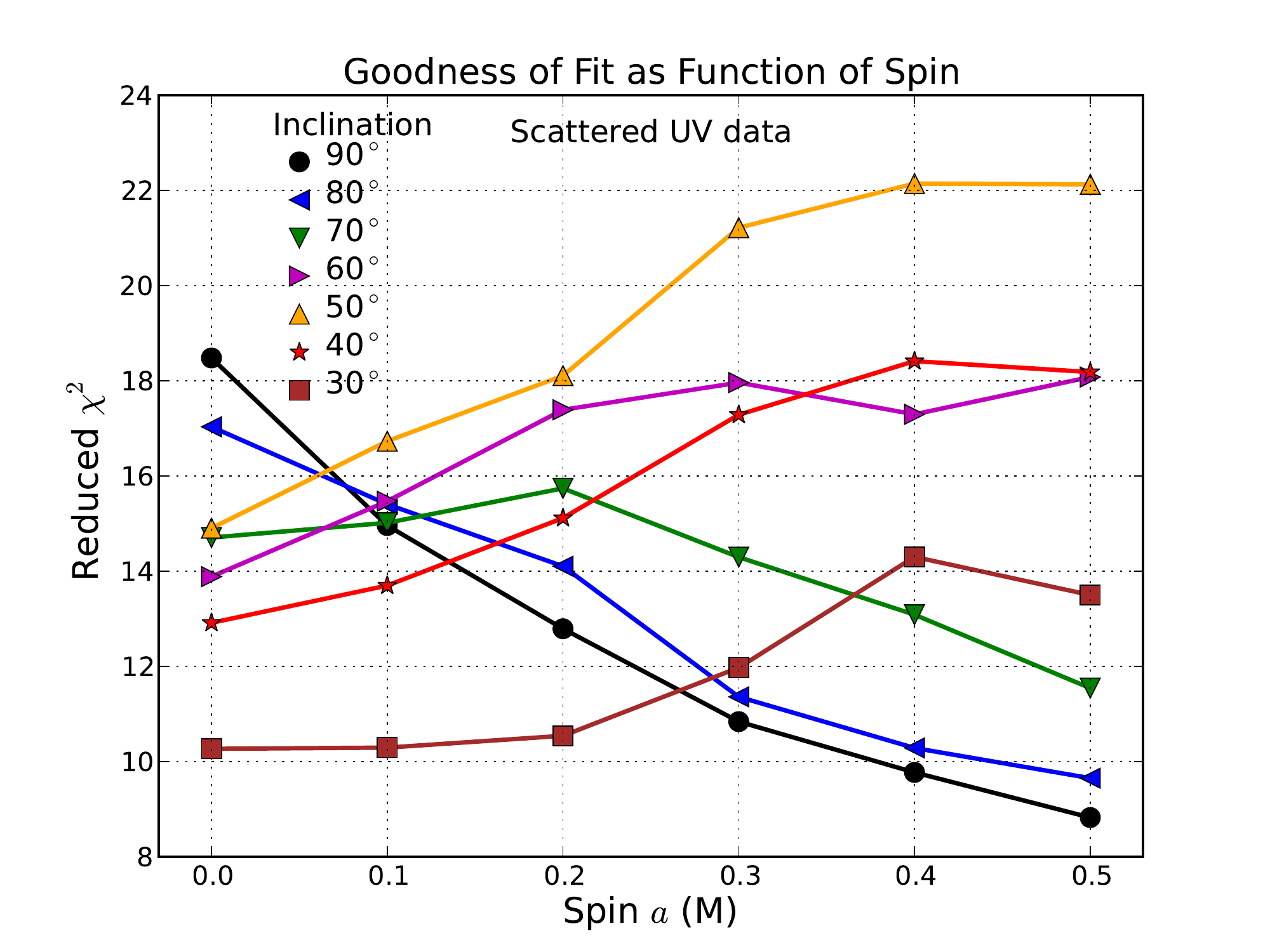}{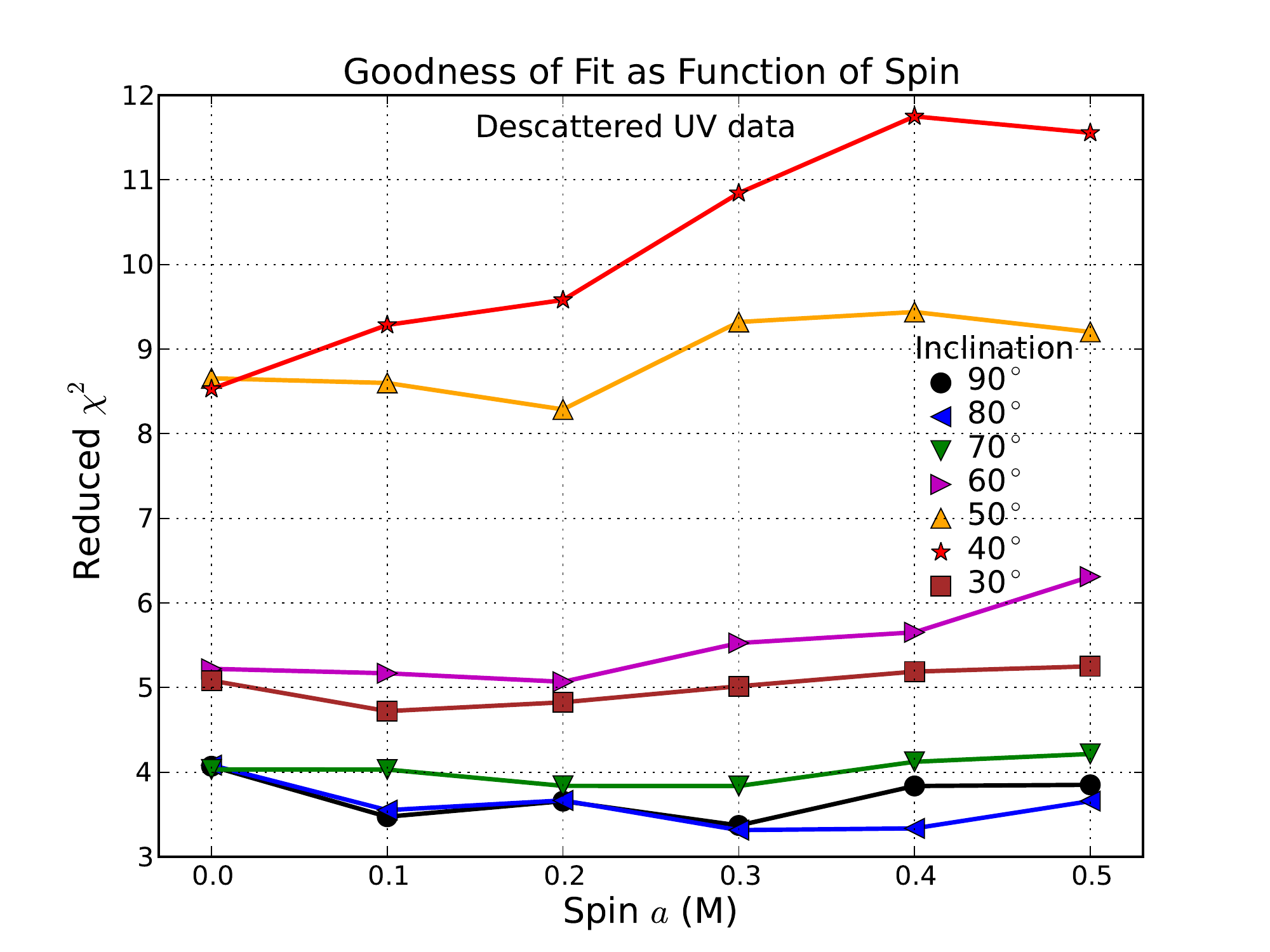}
\caption{\small Dependence of the 9-parameter model fit goodness as reduced $\chi^2$ on the black hole spin for various inclinations.
\label{fig_chi2_vs_spin_snd}}
\end{figure*}

\begin{figure*}[ht!]
\plottwo{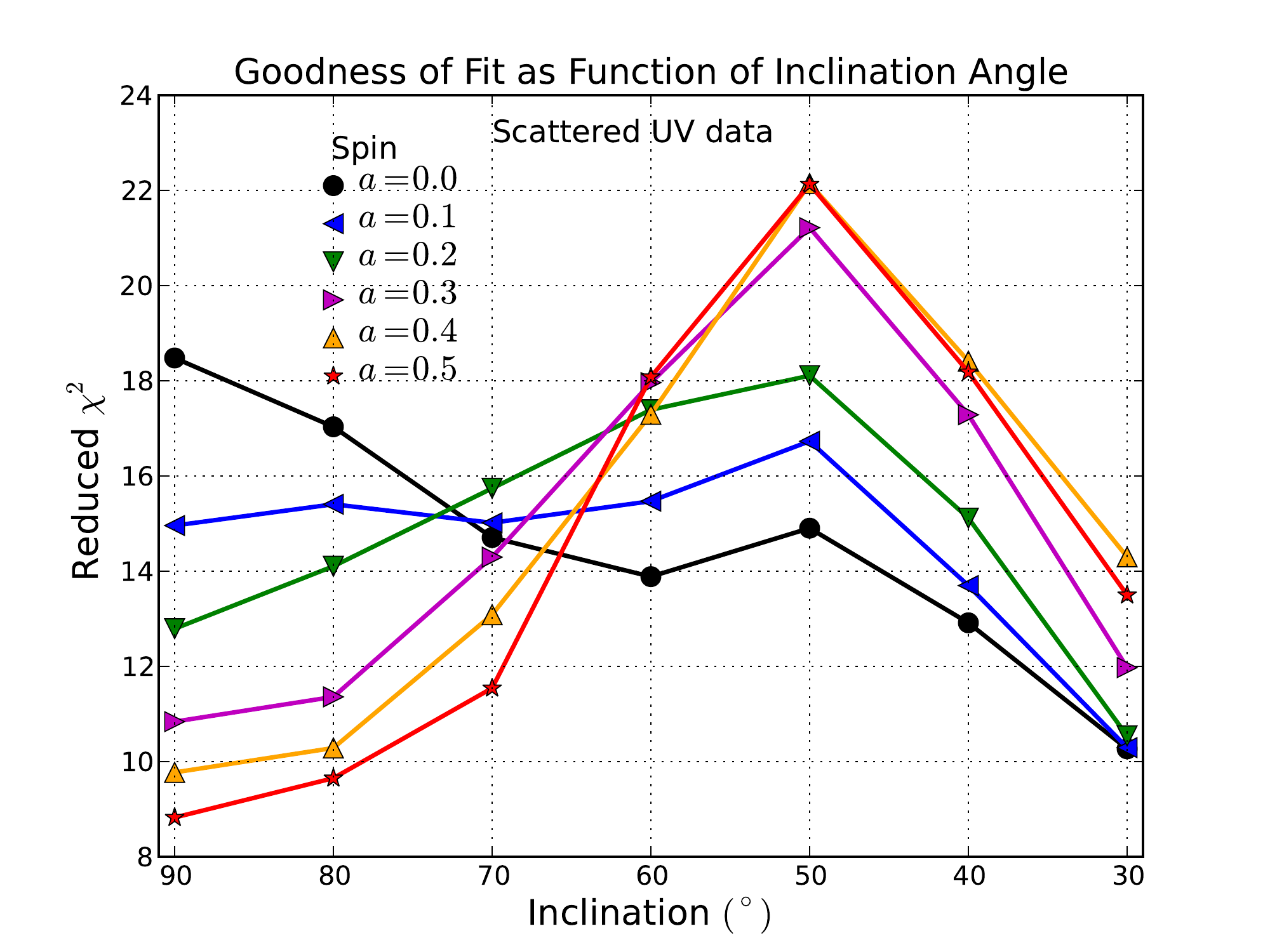}{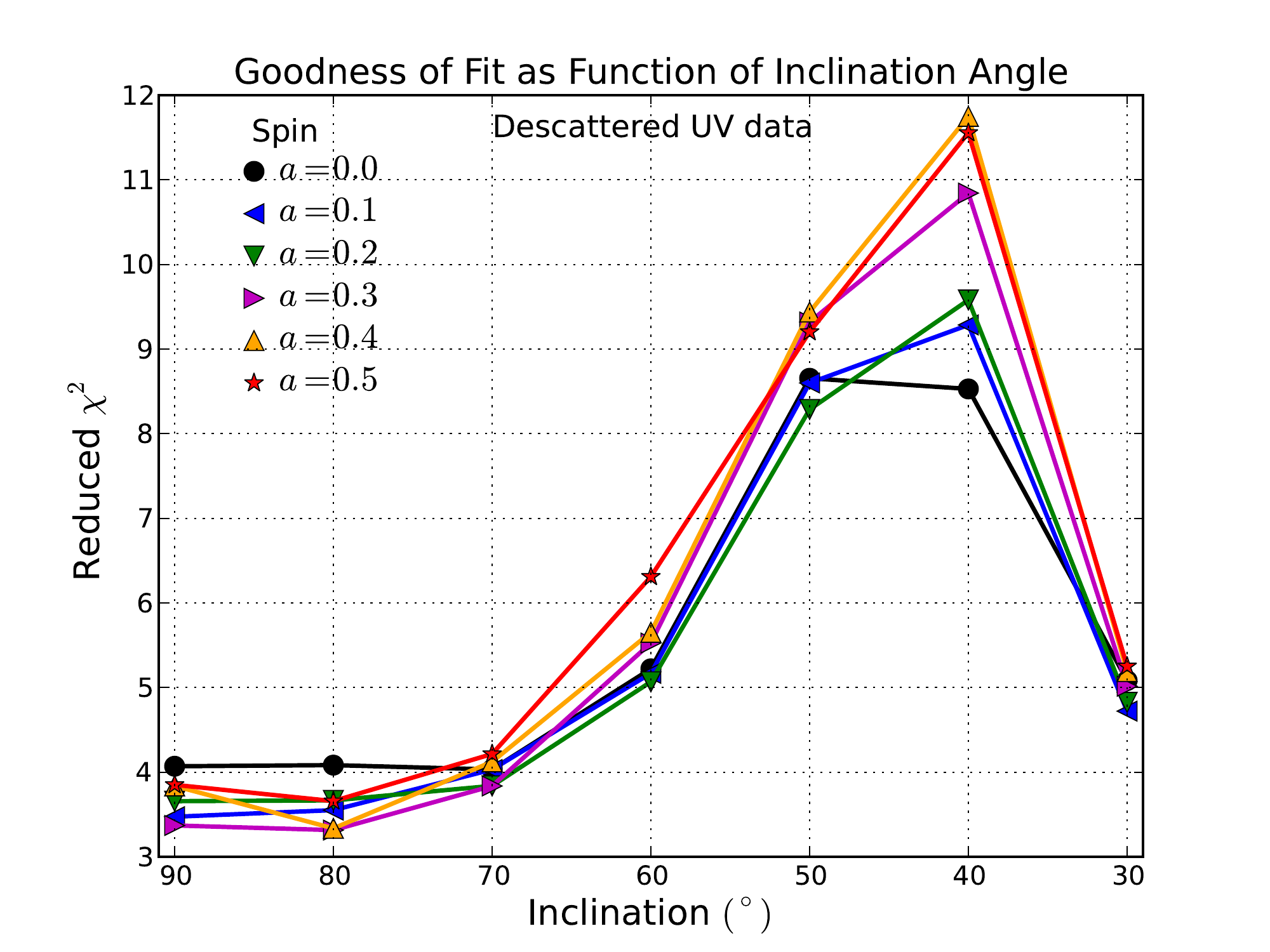}
\caption{\small Dependence of the 9-parameter model fit goodness as reduced $\chi^2$ on the black hole inclination for different spins.
\label{fig_chi2_vs_incl_snd}}
\end{figure*}

Goodness of the model fit is estimated as the reduced $\chi^2$,  
\begin{equation}
  \label{reduced_chi2}
  \chi^2_\nu = \frac{1}{\nu} \chi^2,
\end{equation} 
where $\chi^2$ is calculated by formula \eqref{chi2_distr} and $\nu$ is the number of degrees of freedom calculated as
\begin{equation}
  \label{deg_fre}
  \nu = N_\mathrm{vis} + N_\mathrm{clp} - N_\mathrm{prm} - 1,
\end{equation} 
where $N_\mathrm{vis}$ is the number observed visibilities, $N_\mathrm{clp}$ is the number of closure phases, and $N_\mathrm{prm} = 9$ is the number of fitted model parameters. The fit goodness is dependent on the spin and the inclination. Fig.~\ref{fig_chi2_vs_spin_snd} shows dependences of $\chi^2_\nu$ on the spin for all possible inclinations from $30^\circ$ to $90^\circ$. The right panel is for the model fits to the raw data, while the right panel is for the fits to the descattered data. Descattering significantly improves the goodness of model fits (notice the vertical scale difference). Also, it shows that $\chi^2_\nu$ grows with the decreasing inclination angle. Conversely, Fig.~\ref{fig_chi2_vs_incl_snd} shows dependences of $\chi^2_\nu$ on the inclination for the spins from 0.0 to 0.5 M. Descattering improves the fit goodness, and in the right panel one can see that the fit quality is approximately the same for any spin, but it strongly depends on the inclination. The worst fit quality is near 40$^\circ$ - 50$^\circ$, as was illustrated earlier in Fig.~\ref{fig_compar_incl}.

\section{Discussion}

The described \emph{xringaus} (or 9-parameter) model is an intensional simplification of a real black hole accretion image. However, it can provide valuable information on the most general parameters such as the black hole shadow size, its relative position, the spin axis inclination, differences between the brightest and the dimmest parts etc. The \emph{xringaus} model is a development of the \emph{crescent} model independently designed by \citet{Kamruddin_Dexter_2013}. The \emph{xringaus} model provides a more detailed and hence more informative image. One of the interesting properties of the 9-parameter model is that it is capable to significantly eliminate the effects of interstellar scattering.

We chose modeling in the visibility domain mostly for computational speed. A model in the brightness domain would impose an overhead of a large number of fast Fourier transforms (FFTs) during the MCMC fitting process for every variation of the model parameters. However, modeling in the brightness domain could provide greater flexibility: we would not be restricted to the circular pillboxes and Gaussians. Instead, it would be possible to use any conceivable mathematical forms, non-circular and asymmetric shapes. For example, some authors \citep{deVries2000,Vries2005limacon,Cruz_etal2011,Villanueva_etal2013} consider the Durer-Pascal lima\c{c}on as the mathematical curve describing the shadow. Suppose a parametric image with a non-circular shadow is specified in the brightness domain. Note that the $\chi^2$ computation does not require the Fourier transform of the whole $N\times N$  brightness image. With a moderate number of observational data points, the direct discrete Fourier transform (DFT) of the model brightness into the visibility for only those particular points can be an order of magnitude faster than the FFT producing the whole visibility image. Therefore, the next step in this work is envisioned as Sgr A* image modeling in the brightness domain.

In this numerical study we assumed slow variation of the black hole object, such that it can be considered static over the full track of the observations (over 24 hours). However, Sgr A* is highly variable on a time scale of minutes. M.~Moscibrodzka and J.~Dolence \citep{Monika2012galacenter} developed GRMHD and RIAF models of the black hole accretion flow. Their simulation results in the form of 24 hour Sgr A* ``movies" with the frames only 10 s apart, providing valuable material for future testing of our model-fitting approach on the dynamic images.

\clearpage

\bibliographystyle{apalike}

\bibliography{blackhole_image_reconstr_arxiv_org}
\end{document}